\documentclass[11pt,a4paper]{article}

\usepackage{jheppub,bm}
\usepackage{jheppub}

\usepackage{amsmath,amssymb}
\usepackage{xcolor}
\usepackage[T1]{fontenc}
\usepackage{tocloft}
\usepackage[normalem]{ulem}
\usepackage{subcaption}
\usepackage{array}
\usepackage{enumitem}
\usepackage{extarrows}
\usepackage{mathtools}
\usepackage{slashed} 
\usepackage{physics} 
\usepackage{autobreak}

\allowdisplaybreaks[1] 

\newcommand{\ptv}{p_T^{\rm veto}}

\newcommand{\eps}{\epsilon}

\newcommand{\no}{\nonumber}
\newcommand{\omg}{\omega}

\newcommand{\als}{\alpha_s}

\newcommand{\widebar}[1]{\,\overline{\!{#1}}}
\newcommand{\grid}{{\rm Grid}}
\newcommand{\calC}{{\cal C}}

\newcommand{\df}{{\rm d}}
\newcolumntype{P}[1]{>{\centering\arraybackslash}p{#1}}
\makeatletter
\def\@fpheader{~}
\makeatother

\title{The NNLO gluon beam function for jet-veto resummation}

\author[a]{Guido Bell,} 
\author[a]{Kevin Brune,} 
\author[b]{Goutam Das,} 
\author[c,d]{Ding Yu Shao,} 
\author[a]{and Marcel Wald}

\note{SI-HEP-2024-02, TTK-24-01, P3H-24-004}

\affiliation[a]{
Theoretische Physik 1, Center for Particle Physics Siegen,\\ 
Universit\"at Siegen, 57068 Siegen, Germany}

\affiliation[b]{
Institut f\"ur Theoretische Teilchenphysik und Kosmologie, \\
RWTH Aachen University, 52056 Aachen,
Germany}

\affiliation[c]{
Department of Physics, Center for Field Theory and Particle Physics, \\
and Key Laboratory of Nuclear Physics and Ion-beam Application (MOE) \\
Fudan University, Shanghai 200438, China}

\affiliation[d]{Shanghai Research Center for Theoretical Nuclear Physics, \\
NSFC and Fudan University, Shanghai 200438, China}

\emailAdd{bell@physik.uni-siegen.de}
\emailAdd{brune@physik.uni-siegen.de}
\emailAdd{goutam@physik.rwth-aachen.de}
\emailAdd{dingyu.shao@cern.ch}
\emailAdd{marcel.wald@uni-siegen.de}

\abstract{
We compute the gluon beam function for jet-veto resummation to next-to-next-to-leading order (NNLO) in the strong-coupling expansion. Our calculation is based on an automated framework that was previously used for the computation of the respective quark beam function, and which we significantly extended for the present calculation. In particular, the perturbative matching kernels are directly calculated in momentum space, without the need to perform an additional Mellin transform. We present results for both gluon and quark-initiated processes, which we cross-checked with an independent semi-analytical method that exploits the similarity of the beam functions to the more familiar case of transverse-momentum resummation. Our computation is relevant for jet-veto resummations at NNLL$'$ accuracy.  
}


\begin{document} 

\maketitle

\flushbottom

%%%%%%%%%%%%%%%%%%%%%%%%%%%%%%%%%%%%%%%%%%%
\section{Introduction}
%%%%%%%%%%%%%%%%%%%%%%%%%%%%%%%%%%%%%%%%%%%

Many experimental analyses at the Large Hadron Collider (LHC) apply jet vetoes to enhance the signal sensitivity. In these analyses, the jet veto is usually imposed as a cut on the transverse momenta $\ptv$ of the reconstructed jets. If this scale is much smaller than the typical hard scale  $Q$ of the process, the QCD radiation gets constrained to be either soft or collinear to the beam directions, which induces large Sudakov-type corrections of the form $\alpha_S^n \ln^{2n} (\ptv/Q)$. These logarithms challenge the convergence of the perturbative expansion, and they need to be resummed to all orders to obtain reliable predictions.

The resummation of jet-veto logarithms has received considerable attention in recent years. While the formalism was originally developed for Higgs-boson and Drell-Yan production~\cite{Banfi:2012yh,Becher:2012qa,Tackmann:2012bt,Banfi:2012jm,Becher:2013xia,Stewart:2013faa,Banfi:2015pju,Monni:2019yyr}, it was later applied to processes with other electroweak final states~\cite{Shao:2013uba,Li:2014ria,Jaiswal:2014yba,Becher:2014aya,Wang:2015mvz,Dawson:2016ysj,Campbell:2023cha,Gavardi:2023aco} and beyond-the-standard-model signatures~\cite{Tackmann:2016jyb,Ebert:2016idf,Fuks:2017vtl,Arpino:2019fmo}. In most of these analyses, the jet-veto logarithms were resummed to next-to-next-to-leading logarithmic (NNLL) accuracy, but the theoretical predictions can be further improved using methods from effective field theory. In Soft-Collinear Effective Theory (SCET)~\cite{Bauer:2000yr,Bauer:2001yt,Beneke:2002ph} the effects from the relevant scales $\ptv\ll Q$ are systematically disentangled (factorised), and the logarithmic corrections are resummed using renormalisation-group (RG) techniques. The case of a $\ptv$ veto is special in the sense that it contains rapidity logarithms, which cannot be resummed with conventional RG methods. One therefore resorts to collinear-anomaly~\cite{Becher:2010tm,Becher:2011pf} or rapidity RG techniques~\cite{Chiu:2012ir} for these SCET-2 type observables. 

The starting point of the SCET analyses is a factorisation theorem, which at leading power in $\ptv/Q\ll 1$ can be written in the schematic form 
\begin{align}
\label{eq:factorization}
\frac{\df^2\sigma(\ptv)}{\df Q^2\df Y} = \sum_{i,j} \, H_{ij}(Q,\mu) \, \mathcal{B}_{i/h_1}(x_1,\ptv,\mu) \, \mathcal{B}_{j/h_2}(x_2,\ptv,\mu) \,\mathcal{S}_{ij}(\ptv,\mu)\,,
\end{align}
where the sum runs over all partonic channels, $Q$ is the invariant mass and $Y$ the rapidity of the colourless final state with $x_{1,2}= (Q/\sqrt{s})\,e^{\pm Y}$. Here the purely virtual corrections to the Born process are contained in the hard functions $H_{ij}$, whereas the beam functions $\mathcal{B}_{i/h}$ and the soft functions $\mathcal{S}_{ij}$ describe the effects from collinear and soft radiation, respectively. The factorisation theorem furthermore assumes that the jet radius $R$ satisfies \mbox{$\lambda \ll R \ll \ln (1/\lambda)$} with $\lambda=\ptv/Q$~\cite{Becher:2012qa}.

While the hard functions $H_{ij}(Q,\mu)$ are process-dependent, the soft functions $\mathcal{S}_{ij}(\ptv\!\!,\mu)$ trivially depend on the process via the colour representations of the partons $i,j$. They are currently known to next-to-next-to-leading order (NNLO) accuracy~\cite{Stewart:2013faa,Bell:2020yzz,Abreu:2022sdc}. The beam functions $\mathcal{B}_{i/h}(x,\ptv,\mu)$, on the other hand, are universal and depend on the \mbox{parton $i$} within the hadron $h$ that carries the longitudinal momentum fraction $x$ after emitting collinear radiation that passes the jet-veto constraint. They are by themselves non-perturba\-tive objects, but as long as the jet-veto scale satisfies \mbox{$\ptv\gg\Lambda_{\rm QCD}$}, they can be matched onto the standard parton distribution functions as we will review below. The corresponding matching kernels can then be computed perturbatively, and they were determined to NLO in~\cite{Becher:2012qa,Shao:2013uba}. A subset of the current authors extended this calculation to NNLO for quark-initiated processes~\cite{Bell:2022nrj}, whereas the full set of matching kernels was determined afterwards in~\cite{Abreu:2022zgo}. These results were subsequently used to perform NNLL$'$ resummations in \cite{Campbell:2023cha,Gavardi:2023aco}.\footnote{In the primed-order counting the matching corrections are included at one order higher than in the unprimed one, i.e.~at $\mathcal{O}(\alpha_s^2)$ at NNLL$'$ accuracy.}

The purpose of this work is twofold. First, we will complete the NNLO calculation initiated in~\cite{Bell:2022nrj} by computing the matching kernels for gluon-initiated processes. This serves, on the one hand, as a cross-check of the calculation in~\cite{Abreu:2022zgo}, which was based on a different technique and, in particular, a different prescription to regularise rapidity divergences. We will also study the endpoint behaviour of the matching kernels in some detail since these are often divergent and therefore hard to control numerically. In addition we for the first time present results for the gluon beam function in Mellin space. Second, we stress that the method we use for the calculation of the jet-veto kernels is generic, and it can be applied to a much broader class of observables. It actually follows a strategy that was successfully applied for the calculation of soft functions within the {\tt SoftSERVE} approach~\cite{Bell:2018vaa,Bell:2018oqa,Bell:2020yzz,Bell:2023yso}, which relies on a universal parametrisation of the observable-dependent measurement function in Laplace space. In our previous study~\cite{Bell:2022nrj}, we performed a Mellin transform to resolve the distributions associated with the momentum fraction $x$, whereas we extract these distributions directly in momentum space in the current work. We view this extension as a major improvement of our automated approach, and we anticipate many further results of this novel framework in the future. We, in fact, already recalculated the matching kernels for transverse-momentum resummation and the event-shape variable N-jettiness in this setup~\cite{Bell:2021dpb,Bell:2022tmi,Wald:thesis}.

To validate our results, we performed an independent semi-analytical calculation that is closer to the method that was used in~\cite{Abreu:2022zgo}. Specifically, it exploits the similarity of the jet-veto beam functions to the ones relevant for transverse-momentum resummation, which are known analytically to the considered NNLO and beyond~\cite{Catani:2013tia,Gehrmann:2014yya,Luo:2019szz,Ebert:2020yqt,Luo:2020epw}. In the difference between the two beam functions, many contributions drop out and the calculation can moreover directly be performed in four space-time dimensions. While similar in spirit, our method is not equivalent to the one that was used in~\cite{Abreu:2022zgo} as we will explain below.

The remainder of the paper develops as follows: We first introduce the theoretical foundation of our framework in Sec.~\ref{sec:framework}. In Sec.~\ref{sec:computation} we provide the computational details of both our automated numerical approach and the semi-analytical method that was used for cross-checks. In Sec.~\ref{sec:results} we present our results for the gluon beam function as well as a comparison to the results from \cite{Abreu:2022zgo}. We conclude in Sec.~\ref{sec:conclusions}, collect the relevant anomalous dimensions and splitting functions in App.~\ref{app:anoD}, and provide details on the reference observable that was used in the semi-analytical approach in App.~\ref{app:reference}. Finally, in App.~\ref{app:quark-bf} we present our novel results for the quark beam function in momentum space.

%%%%%%%%%%%%%%%%%%%%%%%%%%%%%%%%%%%%%%%%%%%
\section{Theoretical framework}
%%%%%%%%%%%%%%%%%%%%%%%%%%%%%%%%%%%%%%%%%%%
\label{sec:framework}

To set up the notation, we introduce two light-like vectors $n^{\mu}$ and $\bar{n}^{\mu}$ that obey $n^2=\bar{n}^2=0$ and $n \cdot \bar{n}=2$. Any four-vector $k^\mu$ can then be decomposed according to its projections $k^-=\bar{n} \cdot k$, $k^+=n \cdot k$, and a transverse component $k^{\perp}_\mu$ that satisfies $n \cdot k^{\perp} = \bar{n} \cdot k^{\perp}=0$. In this notation, the gluon beam function for jet-veto resummation is defined as
\begin{align}
\label{eq:definition}
    {\cal B}_{g/h}(x,\ptv,\mu) =&  -(x P^-) \sum_{X} \,
    \delta\Big( (1-x) P^- - \sum_i k_i^- \Big)\,
		\widehat{\mathcal{M}}(\ptv;\{k_i\})
		\nonumber\\
    & \times \bra{h(P)}\mathcal{A}_{n\perp}^{c,\mu}(0) \ket{X} 
    \bra{X} \mathcal{A}_{n\perp,\mu}^{c}(0)  \ket{h(P)} ,
\end{align}
where $\mathcal{A}_{n \perp}^{c,\mu}(x)=1/g_s \, W_{n}^{\dagger}(x)\big[i D^{c,\mu}_{\perp} \,W_{n} (x)\big]$ is the $n$-collinear gluon field operator, $W_{n}$ a collinear Wilson line, and the sum over $X$ represents the phase space of the final-state partons with momenta $\{k_i\}$. At the tree level, when there is no collinear emission, this is simply the vacuum state, and at NNLO it denotes the phase space of up to two massless partons. The hadronic state with collinear momentum $P^{\mu}=P^- n^{\mu}/2$ is furthermore indicated by $\ket{h(P)}$, whereas the jet-veto constraint is imposed by the measurement function $\widehat{\mathcal{M}}(\ptv;\{k_i\})$, which we will specify in Sec.~\ref{sec:computation} below. 

As a matrix element of hadronic states, the beam function is not directly accessible in perturbation theory. But as long as the scale that constrains the collinear emissions satisfies $\ptv\gg\Lambda_{\rm QCD}$, it can be matched onto the usual parton distribution functions $f_{i/h}(x,\mu)$ according to
\begin{align}
\label{eq:matching}
\mathcal{B}_{i/h}(x,\ptv,\mu) &=\sum_k 
    \int_x^1 \frac{\df z}{z} \,
    {\cal I}_{i\leftarrow k}(z,\ptv,\mu) 
    ~ f_{k/h}\Big(\frac{x}{z},\mu\Big)\,,
\end{align}
which holds at leading power in $\Lambda_{\rm QCD}/\ptv$.
Our goal consists in computing the matching kernels ${\cal I}_{i\leftarrow k}(z,\ptv,\mu)$ for gluon-initiated processes ($i=g$) to NNLO in QCD. However, since our previous results for the quark channels ($i=q$) were provided only in Mellin space~\cite{Bell:2022nrj}, we will for completeness also determine those kernels directly in momentum space in this work. The matching calculation can, in fact, be most easily performed using partonic on-shell states, since the parton distribution functions then evaluate to \mbox{$f_{i/j}(x,\mu)=\delta_{ij}\,\delta(1-x)$} to all orders in perturbation theory when dimensional regularisation is used to regularise both ultraviolet (UV) and infrared (IR) divergences. The partonic calculation therefore directly yields the matching kernels in this case.

As stated in the introduction, the jet veto is usually imposed on the transverse momenta of the reconstructed jets. It is well known, however, that transverse-momentum dependent observables belong to a special class of observables that suffer from rapidity divergences that are not captured by the dimensional regulator $\eps=(4-d)/2$. This follows from the fact that the relevant soft and collinear modes have the same virtuality in the effective theory, which is also known as SCET-2. To regularise these rapidity divergences, we use the analytic phase-space regulator from~\cite{Becher:2011dz}, which is introduced for each emission with momentum $k_i^\mu$ via
\begin{align}
\label{eq:regulator}
\int \df^dk_i \; \left(\frac{\nu}{k_i^- + k_i^+}\right)^\alpha \;  \delta(k_i^2) \, \theta(k_i^0) \,,
\end{align}
where $\alpha$ is the rapidity regulator and $\nu$ the associated rapidity scale. The same prescription is also implemented for the soft integrals in {\tt SoftSERVE}, but for collinear emissions with $k_i^- \gg k_i^+$ it can be further simplified at leading power. The very fact that the rapidity regulator in \eqref{eq:regulator} respects the $n$-$\bar{n}$ symmetry of the process, ensures that the beam functions for collinear and anti-collinear emissions are equivalent to all orders in perturbation theory.

The matching kernels defined in \eqref{eq:matching} thus depend on the scheme that is used to regularise the rapidity divergences and, in particular, the rapidity scale $\nu$. To obtain a result that is scheme independent, we follow the collinear-anomaly approach~\cite{Becher:2010tm,Becher:2011pf}, which states that the product of the soft and  collinear functions can be refactorised as
\begin{align}
  &\Big[
    {\cal I}_{g\leftarrow i}(z_1,\ptv,\mu,\nu) \;
    {\cal I}_{g\leftarrow j}(z_2,\ptv,\mu,\nu) \; 
		    {\cal S}_{gg}(\ptv,\mu,\nu) 
  \Big]_{Q}
	\nonumber\\
  &\qquad=
  \left( \frac{Q}{\ptv}\right)^{-2F_{gg}(\ptv,\mu)} \;
  I_{g\leftarrow i}(z_1,\ptv,\mu) \;
  I_{g\leftarrow j}(z_2,\ptv,\mu)\,,
	\label{eq:refact}
\end{align}
where the matching kernels $I_{i\leftarrow j}(z,\ptv,\mu) $ on the right-hand side of this equation are independent of the rapidity scale $\nu$. The dependence on the hard scale $Q$ is furthermore resummed through the collinear-anomaly exponent $F_{gg}(\ptv,\mu)$, which satisfies the RG equation 
\begin{align}
\label{eq:anomaly:RGE}
  \frac{\df}{\df\ln \mu} \; F_{gg}(\ptv,\mu) 
  = 2\,\Gamma_{\rm cusp}^A(\als)
\end{align} 
with the cusp anomalous dimension in the adjoint representation $\Gamma_{\rm cusp}^A(\als)$. Up to two-loop order, its solution is given by
\begin{align}
\label{eq:anomaly:RGE:solution}
    F_{gg}(\ptv,\mu)=\left(\frac{\als}{4\pi}\right)\left\{ 2 \Gamma_0^A L +d_1^A\right\}+\left(\frac{\als}{4\pi}\right)^2\left\{ 2 \beta_0\Gamma_0^A L^2 +2\left(\Gamma_1^A +\beta_0 d_1^A\right)L+d_2^A\right\}\, ,
\end{align}
where $L=\ln (\mu/\ptv)$, and $\Gamma_i^A$ and $\beta_i$ are the expansion coefficients of the cusp anomalous dimension and the $\beta$-function, respectively, that are given explicitly to the required order in App.~\ref{app:anoD}. As the anomaly exponent describes the physics from the overlap of the soft and collinear regions, the non-logarithmic terms $d_i^A$ in \eqref{eq:anomaly:RGE:solution} are constrained by Casimir scaling, and they are therefore related to the corresponding quantities for the quark channels, which were given in our previous work \cite{Bell:2022nrj}. We also note that the anomaly exponent renormalises additively, $F_{gg}^{\rm bare}=F_{gg}+Z^F_{gg}$, and the corresponding $\widebar{\text{MS}}$ counterterm $Z^F_{gg}$ obeys a similar RG equation as in \eqref{eq:anomaly:RGE}, whose two-loop solution reads
\begin{align}
    Z^F_{gg}(\ptv,\mu)=\left(\frac{\als}{4\pi}\right)\left\{ \frac{\Gamma_0^A}{\eps}\right\}+\left(\frac{\als}{4\pi}\right)^2\left\{-\frac{\beta_0\Gamma_0^A}{2\eps^2}+\frac{\Gamma_1^A}{2\eps}\right\}\,.
\end{align}
The remaining ingredients in the collinear-anomaly relation \eqref{eq:refact} are the refactorised matching kernels $I_{i\leftarrow j}(z,\ptv,\mu)$, which are independent of the rapidity regularisation scheme. Their scale dependence is determined by the RG equation
\begin{align}
\label{eq:rgeI}
  \frac{\df}{\df\ln \mu} \;I_{i\leftarrow j}(z,\ptv,\mu)
  &=
  2\left[
    \Gamma_{\rm cusp}^{R_i}(\als) \, L -\gamma^{i}(\als)
  \right] I_{i\leftarrow j}(z,\ptv,\mu)
	\nonumber\\
  &\quad -2 \sum_{k} \int_z^1 \frac{\df z'}{z'} \,
    I_{i\leftarrow k}(z',\ptv,\mu) 
    ~ P_{k\leftarrow j}\Big(\frac{z}{z'},\alpha_s\Big)\,,
\end{align}
where $\Gamma_{\rm cusp}^{R_i}(\als)$ is the cusp anomalous dimension in the representation of the parton $i$, $\gamma^{i}(\als)$ is the collinear anomalous dimension for quarks $(i=q)$ or gluons $(i=g)$, and $P_{k\leftarrow j}(z,\als)$ are the DGLAP splitting functions. The renormalisation of the refactorised matching kernels is, in fact, slightly more complicated, and it is convenient to introduce two types of counterterms that subtract the UV divergences of the beam function ($Z_i^B$) and the IR divergences associated with the matching onto the parton distribution functions ($Z_{k\leftarrow j}^f$) for them~\cite{Bell:2022nrj}. Specifically, we write 
\begin{align}
    {I}_{i\leftarrow j}(z,\ptv,\mu) = Z_i^B(\ptv,\mu)\, \sum_k\int_z^1 \frac{\df z'}{z'} \,
    I_{i\leftarrow k}^{\rm bare}(z',\ptv) 
    ~ Z_{k\leftarrow j}^f\Big(\frac{z}{z'},\mu\Big)\,,
\end{align}
where the UV counterterm obeys the RG equation
\begin{align}
  \frac{\df}{\df\ln \mu} \;
	Z_i^B(\ptv,\mu)
  =&
  2\left[
    \Gamma_{\rm cusp}^{R_i}(\als) \, L -\gamma^{i}(\als)
  \right] Z_i^B(\ptv,\mu) \,.
\end{align}
Up to two loops, its solution reads
\begin{align}
&Z_i^B(\ptv,\mu)  =  
1 + \left( \frac{\alpha_s}{4 \pi} \right) 
\left\{ - \frac{\Gamma_0^i}{2\eps^2} - 
\frac{\Gamma_0^i L - \gamma_0^i}{\eps} 
\right\}
\nonumber\\[0.2em]  
 &\quad
+\left( \frac{\alpha_s}{4 \pi} \right)^2 
\bigg\{ \frac{(\Gamma_0^i)^2}{8\eps^4}
+ \left( \frac{\Gamma_0^i}{2}L
-\frac{\gamma_0^i}{2}
 + \frac{3\beta_0}{8}\right) \frac{\Gamma_0^i}{\eps^3}
+\bigg( \frac{(\Gamma_0^i)^2}{2}L^2 
- \Gamma_0^i \Big(
\gamma_0^i - \frac{\beta_0}{2} \Big)L
\nonumber\\[0.2em]  
 &\qquad\qquad\qquad
-\frac{\Gamma_1^i}{8} 
+ \frac{(\gamma_0^i)^2}{2}
- \frac{\beta_0\gamma_0^i}{2}
\bigg) \, \frac{1}{\eps^2}
 - \frac{\Gamma_1^i L - \gamma_1^i}{2\eps}
\bigg\}
\,,
\end{align}
and the specific values for the coefficients of the anomalous dimensions are also summarised in App.~\ref{app:anoD}. The IR counterterm, on the other hand, satisfies the RG equation
\begin{align}
  \frac{\df}{\df\ln \mu} \;
	Z_{k\leftarrow j}^f(z,\mu)
  =&
  -2 \sum_{l}  \int_z^1 \frac{\df z'}{z'} \,
    Z_{k\leftarrow l}^f(z',\mu) 
    ~ P_{l\leftarrow j}\Big(\frac{z}{z'},\alpha_s\Big)\,,
\end{align}
whose two-loop solution is given by
\begin{align}
&Z_{k\leftarrow j}^f(z,\mu)  =  
\delta_{kj}\,\delta(1-z) + \left( \frac{\alpha_s}{4 \pi} \right) 
\left\{ P_{k\leftarrow j}^{(0)}(z)\,\frac{1}{\eps}
\right\}
\nonumber\\[0.2em]  
&\quad
+\left( \frac{\alpha_s}{4 \pi} \right)^2 
\bigg\{ 
- P_{k\leftarrow j}^{(0)}(z)\,
\frac{\beta_0}{2\eps^2}
+\frac{1}{2\eps^2} \Big( P_{k\leftarrow l}^{(0)} \otimes
 P_{l\leftarrow j}^{(0)} \Big) (z) 
+ P_{k\leftarrow j}^{(1)}(z)\,\frac{1}{2\eps}\bigg\}
\,,
\label{eq:Zfren}
\end{align}
where the coefficients of the DGLAP splitting functions ${P}_{k\leftarrow j}^{(i)}(z)$ are defined in App.~\ref{app:anoD}. We furthermore introduced a short-hand notation for the convolutions
\begin{align}
    \Big( P_{k\leftarrow l}^{(0)} \otimes
 P_{l\leftarrow j}^{(0)} \Big) (z) &=
 \sum_{l}  \int_z^1 \frac{\df z'}{z'} \,
    P_{k\leftarrow l}^{(0)}(z') 
    ~ P_{l\leftarrow j}^{(0)}\Big(\frac{z}{z'}\Big)
\label{eq:def:convolution}    
\end{align}
that appear frequently in the calculation. Note that this notation implies a sum over intermediate partonic channels with index $l$. For completeness, we give the explicit expressions for these convolutions also in App.~\ref{app:anoD}.

Finally, we state the solution of the RG equation \eqref{eq:rgeI} for the renormalised matching kernels. Up to NNLO, it takes the form
\begin{align}
\label{eq:Iren}
&I_{i\leftarrow j}(z,\ptv,\mu)  
\\[0.2em]  
&= 
\delta_{ij}\,\delta(1-z) + \left( \frac{\alpha_s}{4 \pi} \right) 
\bigg\{ \Big( \Gamma_0^i \,L^2 
- 2\gamma_0^i \,L \Big) \delta_{ij}\,\delta(1-z)  
- 2L\,P_{i\leftarrow j}^{(0)}(z)
+ I_{i\leftarrow j}^{(1)}(z) \bigg\}
\nonumber\\[0.2em]  
  &
+\left( \frac{\alpha_s}{4 \pi} \right)^2 
\bigg\{ \bigg( \frac{(\Gamma_0^i)^2}{2} L^4 
-  2\Gamma_0^i\left( \!\gamma_0^i - \frac{\beta_0}{3} \right) 
L^3 + \big( \Gamma_1^i + 2(\gamma_0^i)^2 
 - 2\beta_0 \gamma_0^i  \big) L^2  
-  2\gamma_1^i L\bigg) \delta_{ij}\,\delta(1-z)
\nonumber\\[0.2em]  
 &\qquad\qquad\quad
-2\Big( \Gamma_0^i L^3 + \big( \beta_0 - 2 \gamma_0^i\big) L^2\Big) P_{i\leftarrow j}^{(0)}(z)
 + \Big( \Gamma_0^i  L^2 -  2(\gamma_0^i -\beta_0) L \Big)\,
I_{i\leftarrow j}^{(1)}(z)  
\nonumber\\[0.2em]  
 &\qquad\qquad\quad
 +2 L^2 \Big( P_{i\leftarrow k}^{(0)} \otimes P_{k\leftarrow j}^{(0)}  \Big)(z)
-2 L \Big( I_{i\leftarrow k}^{(1)} \otimes  
P_{k\leftarrow j}^{(0)} \Big)(z)  
- 2L\,P_{i\leftarrow j}^{(1)}(z)
+  I_{i\leftarrow j}^{(2)}(z) \bigg\} .
\nonumber
\end{align}
As the logarithmic terms in this expression are already known, we will focus on the non-logarithmic terms $I_{i\leftarrow j}^{(m)}(z)$ in later sections. 

%%%%%%%%%%%%%%%%%%%%%%%%%%%%%%%%%%%%%%%%%%%
\section{Computational aspects}
%%%%%%%%%%%%%%%%%%%%%%%%%%%%%%%%%%%%%%%%%%%
\label{sec:computation}

As stated in the introduction, we followed two different approaches to compute the renormalised matching kernels for jet-veto resummation. The first approach is based on an automated numerical framework that can be used for a generic class of observables, while the second approach is a semi-analytical method that exploits the similarity of the jet-veto beam functions to the ones relevant for transverse-momentum resummation. We will now describe each of these methods in turn.

%%%%%%%%%%%%%%%%%%%%%%%%%
\subsection{Numerical approach}
%%%%%%%%%%%%%%%%%%%%%%%%%
\label{sec:numerical}

In the first approach, we start from the definition of the beam function \eqref{eq:definition} with the measurement function replaced by the ansatz
\begin{align}
\label{eq:measure}
\mathcal{M}(\tau;\lbrace k_{i} \rbrace) = 
\exp\big(-\tau\, \omega(\lbrace k_{i} \rbrace)\,\big)\,,
\end{align}
where $\tau$ is a Laplace variable of dimension 1/mass, and the function $\omega(\lbrace k_{i} \rbrace)$ specifies a generic observable. Following \cite{Bell:2022tmi, Bell:2022nrj}, we find it convenient to work in Laplace space, since both the calculation of the bare beam function and its renormalisation simplify in this case. We note that for the specific jet-veto beam function defined in \eqref{eq:definition}, the measurement function is \emph{not} of the form \eqref{eq:measure}, but it is instead given by \mbox{$\widehat{\mathcal{M}}(\ptv;\lbrace k_{i} \rbrace) = \theta\big(\ptv- \omega(\lbrace k_{i} \rbrace)\,\big)$}. We then follow the procedure described in \cite{Bell:2020yzz} to convert this into the form \eqref{eq:measure} by taking a Laplace transform, and we extract the results in the original $\ptv$ space by inverting the Laplace transformation at the very end of the calculation.

To factorise the implicit divergences of the collinear matrix elements, we use universal phase-space parametrisations as in~\cite{Bell:2022tmi, Bell:2022nrj}.  Particularly, for a single emission with momentum $k^\mu$, we use the magnitude of its transverse momentum relative to the beam axis, which is set by $P^{\mu}\propto n^\mu$, and we allow for a non-trivial azimuthal dependence of the observable, 
\begin{align}
\label{eq:parametrisation:one}
  k_T = |\vec{k}^\perp| \,, 
	\qquad\qquad t_k = \frac{1-\cos \theta_k}{2} \,,
\end{align} 
where $\theta_k$ is the angle between $\vec{k}^\perp$ and a reference vector $\vec v^\perp$ that could be imposed by the observable. The remaining momentum components are then fixed by the on-shell condition and the explicit delta function in (\ref{eq:definition}), yielding $k^+=k_T^2 / k^-$ and $k^-= \bar z P^-$ with $\bar z =1-z$.  This leads to the following measurement function for a single emission,
\begin{align}
\label{eq:measure:ptv:nlo}
\omg_1(k)  = k_T\,.
\end{align}
For two emissions, we proceed similarly and parametrise the momenta $k^\mu$ and $l^\mu$ of the massless final-state partons in the form\footnote{Note that we use a slightly different notation here compared to our setup that was used for the calculation of the Mellin-space kernels~\cite{Bell:2022nrj}.}
\begin{align}
\label{eq:parametrisation:two}
  a = \frac{k^- l_T}{l^- k_T}, \qquad
  b = \frac{k_T}{l_T}, \qquad
  \bar z = \frac{k^- + l^-}{P^-}, \qquad
  q_T = \sqrt{(k^- + l^-)(k^+ + l^+)}\,,
\end{align}
where again $k_T = |\vec{k}^\perp|$, $l_T = |\vec{l}^\perp|$, and we in addition now have three non-trivial angles in the transverse plane, with $\theta_k$ and $\theta_l$ referring to the external vector $\vec{v}^\perp$ as before, and $\theta_{kl}$ being the angle between $\vec{k}^\perp$ and $\vec{l}^\perp$. We then rewrite these angles in terms of variables $t_k$, $t_l$, and $t_{kl}$ that are defined on the unit interval, similar to \eqref{eq:parametrisation:one}.

In terms of these variables, the two-emission measurement function for jet-veto resummation takes the form
\begin{align}
\label{eq:omega2}
 & \omg_2(k,l)   =
\frac{q_T\,\sqrt{a}}{\sqrt{(1+a b)(a+b)}}\,\bigg\{
	\theta(\Delta-R)\; \max(1,b) + \theta(R-\Delta )\;
	\sqrt{(1 - b)^2 + 4 b (1 - t_{kl})} \bigg\}\,,
\end{align}
where $R$ is the jet radius  and  $\Delta= \sqrt{\ln^2 a+\arccos^2(1-2t_{kl})}$ is the distance measure of the jet algorithm. The twofold structure in \eqref{eq:omega2} arises as follows: If the distance between the two emissions is smaller than $R$, the emissions are clustered together and the veto is imposed on the recombined pseudo-particle. If, on the other hand, this distance is larger than $R$, the veto constrains both of the reconstructed jets. Similar to our previous work on the quark beam function~\cite{Bell:2022nrj}, we consider the general class of $k_T$-type jet algorithms in this work, for which the expression in \eqref{eq:omega2} turns out to be independent of the specific clustering prescription ($k_T$, Cambridge/Aachen, anti-$k_T$, \ldots)~\cite{Becher:2012qa}. 

Having defined the phase-space parametrisations and the structure of the measurement function, the partonic beam functions can now be computed. The required collinear matrix elements are related to the spin-averaged $d$-dimensional time-like splitting functions~\cite{Ritzmann:2014mka}. Specifically, we used the expressions from~\cite{Kosower:1999rx,Bern:1999ry,Sborlini:2013jba} for the real-virtual contributions, and the ones from~\cite{Campbell:1997hg,Catani:1998nv} for the double real-emission part, and we applied standard crossing relations to convert these to time-like kinematics~\cite{Wald:thesis}. The main task then consists in factorising the implicit phase-space divergences, which allows one to perform the expansion in the two regulators $\alpha$ and $\eps$ directly on the integrand level. While this can easily be achieved in the one-emission case, this is only true in certain cases for the double real emissions in the parametrisation (\ref{eq:parametrisation:two}). We therefore employed a number of further techniques that involve sector-decomposition steps and non-linear transformations to factorise all divergences for this contribution (details can be found in~\cite{Wald:thesis}). 

In comparison to our previous work on the quark beam function~\cite{Bell:2022nrj}, our novel framework is capable of computing the beam functions directly in momentum space. In particular, within the parameterisations \eqref{eq:parametrisation:one} and \eqref{eq:parametrisation:two} it is possible to factorise the divergence associated with the longitudinal momentum fraction $z$. We may then introduce the corresponding distributions via
\begin{align}
(1-z)^{-1-m \alpha}= - \frac{\delta(1-z)}{m \alpha} + \left[\frac{1}{1-z}\right]_+
- m \alpha \left[\frac{\ln (1-z)}{1-z}\right]_+ + \,\ldots
\end{align}
After expanding the integrand first in the rapidity regulator $\alpha$ and subsequently in the dimensional regulator $\eps$, the coefficients in this double expansion can be integrated numerically. The final expressions for the matching kernels then consist of distributions with numerical coefficients and a non-trivial $z$-dependent `grid' contribution that we sample for different values of $z$. To perform these steps, we have implemented our formalism in the publicly available program {\tt pySecDec} \cite{Borowka:2017idc}, and we use the {\tt Vegas} routine of the {\tt Cuba} library \cite{Hahn:2004fe} for the numerical integrations. By following this procedure, one obtains the regulator-dependent matching kernels on the left-hand side of the refactorisation condition \eqref{eq:refact}. In the last step, we then combine these expressions with the corresponding NNLO soft function that is provided by the {\tt SoftSERVE} distribution to extract the $I_{i\leftarrow j}^{(m)}(z)$ part of the refactorised and renormalised matching kernels that are defined in \eqref{eq:Iren}.

%%%%%%%%%%%%%%%%%%%%%%%%
\subsection{Semi-analytical approach}
%%%%%%%%%%%%%%%%%%%%%%%%%%
\label{sec:analytical}

In the second approach, we use the fact that the beam functions for jet-veto resummation are related to the ones for transverse-momentum resummation. As the latter are known analytically at the considered NNLO, one focuses on the difference 
\begin{align}
\label{eq:deltaI}
  \Delta {\cal I}_{i\leftarrow j}(z,\ptv,\mu,\nu)  = 
  {\cal I}_{i\leftarrow j}(z,\ptv,\mu,\nu) 
  - {\cal I}_{i\leftarrow j}^{\text{ref}}(z,\ptv,\mu,\nu)\,, 
\end{align}
where the first term refers to the jet-veto matching kernels that appear on the left-hand side of the refactorisation condition~\eqref{eq:refact}, and the second term indicates the corresponding kernels for a so-called reference observable. Following~\cite{Becher:2013xia}, we choose for the latter the integrated $q_T$-spectrum,
\begin{align}
\label{eq:def:reference}
  \sigma^{\text{ref}} (\ptv) 
  = \int_0^{\ptv} \!\!\!\!\df q_T \; \frac{\df\sigma}{\df q_T}\,,
\end{align}
and we provide details on the derivation of the corresponding refactorised matching kernels $I_{i\leftarrow j}^{\text{ref},(2)}(z)$ in App.~\ref{app:reference}.

At NLO the difference in \eqref{eq:deltaI} vanishes by construction, and at NNLO it only receives contributions from diagrams with double real emissions. The measurement function capturing these contributions can be expressed as
\begin{align}
\label{eq:measure-diff}
  \widehat{\mathcal{M}}_\Delta(\ptv;\{k,l\}) = 
  \theta(\Delta - R) \left[\theta\big(\ptv - |\vec{k}^\perp|\big) 
  \theta\big(\ptv - |\vec{l}^\perp|\big) 
  - \theta\big(\ptv - |\vec{k}^\perp+\vec{l}^\perp|\big)\right],
\end{align}
where the angular separation $\Delta$ between the two emissions is quantified as before.

As has been further argued in~\cite{Becher:2013xia}, there is a price to pay when working with the difference \eqref{eq:deltaI}, i.e.~one has to explicitly compute mixing terms that originate from different sectors of the effective theory. Notably, these contributions are independent of the jet radius $R$ and they emerge solely within the uncorrelated contribution to the amplitude. To manage these contributions effectively, we decompose the double-emission amplitude in the collinear region $\mathcal{A}_{c}^{(2)}$ into its uncorrelated and correlated components,
\begin{align}
  \mathcal{A}_{c}^{(2)}(k,l,{P}) = \mathcal{A}_{c}^{\text{uncor}, \, (2)}(k,l,{P}) + \mathcal{A}_{c}^{\text{cor}, \, (2)}(k,l,{P})\,,
\end{align}
wherein the uncorrelated term is constructed from the product of one-emission collinear and soft amplitudes,
\begin{align}
  \mathcal{A}_{c}^{\text{uncor}, \, (2)} (k,l,{P})= \frac{1}{2} \mathcal{A}_{c}^{(1)}(k,{P}) \mathcal{A}_{s}^{(1)}(l) +  (k \leftrightarrow l)\,. 
\end{align}
Therefore only specific colour structures receive mixing contributions.
The starting point for the evaluation of \eqref{eq:deltaI} is the representation
\begin{align}
    {\Delta {\cal I}_{i\leftarrow j}^{(2)}(z,\ptv,\mu,\nu) } &= 
    \int \!\frac{\mathrm{d}^4 k}{(2\pi)^3} \, \frac{\mathrm{d}^4 l}{(2\pi)^3}  \; 
    \delta(k^2)\theta(k^0) \, \delta(l^2)\theta(l^0) \,
    \delta\Big(k^-+l^--(1-z)P^-\Big) 
     \notag \\[0.5em]
    &\qquad \times \mathcal{A}_{c}^{(2)}(k,l,{P}) \; 
    \widehat{\mathcal{M}}_\Delta(\ptv;\{k,l\}) \,
    \left(\frac{\nu}{k^-}\right)^\alpha \left(\frac{\nu}{l^-}\right)^\alpha\,,
\end{align}
where the rapidity regulator $\alpha$ has been factored in. As the divergences in the dimensional regulator $\eps$ are absent for the measurement \eqref{eq:measure-diff}, one can immediately set $d=4$ in the calculation. To evaluate this contribution we employ the following phase-space parametrisation~\cite{Becher:2013xia},
\begin{align}
    \Delta y = y_k - y_l, \qquad 
    \Delta\phi = \phi_k - \phi_l, \qquad 
    p_T = k_T + l_T, \qquad 
    {\xi} = \frac{k_T}{p_T}\,.
\end{align}
In the computation of the correlated term, we then utilize integration techniques as delineated in \cite{Becher:2013xia}. Initially, we determine the asymptotic behaviour in the limit $R \rightarrow 0$ by expanding the integrands around this limit. Moreover, the integral includes non-singular terms, which can be represented as a power series in $R^2$. To proceed, we first subtract the leading singularities as $\Delta\phi$ and $\Delta y$ approach zero, and we evaluate the coefficients of the power series in $R^2$ numerically.  

The evaluation of the uncorrelated contribution necessitates consideration of the mixing terms as discussed above.  For this part, we rewrite the theta function $\theta(\Delta-R)$ in \eqref{eq:measure-diff} as $1 - \theta(R-\Delta)$, which bifurcates the analysis into $R$-independent and $R$-dependent terms. The $R$-dependent terms are free of mixing contributions, and their computation parallels the treatment of the correlated part. Conversely, for the $R$-independent terms one starts from the measurement function
\begin{align}
  \widehat{\mathcal{M}}_\Delta^{\text{R-indep.}}(\ptv;\{k,l\}) = 
  \theta\big(\ptv - |\vec{k}^\perp|\big) 
  \theta\big(\ptv - |\vec{l}^\perp|\big) 
  - \theta\big(\ptv - |\vec{k}^\perp+\vec{l}^\perp|\big)\,,
\end{align}
and the mixed collinear/soft contribution can in this case be calculated from
\begin{align}
    & {\Delta {\cal I}_{i\leftarrow j}^{(2,cs)}(z,\ptv,\mu,\nu) } = 
    \int \!\frac{\mathrm{d}^4 k}{(2\pi)^3} \, \frac{\mathrm{d}^4 l}{(2\pi)^3}  \; 
    \delta(k^2)\theta(k^0) \, \delta(l^2)\theta(l^0) \,
    \delta\Big(k^--(1-z)P^-\Big) 
     \notag \\[0.5em]
    &\quad \times \frac12 \mathcal{A}_{c}^{(1)}(k,{P})\,\mathcal{A}_{s}^{(1)}(l) \; 
    \widehat{\mathcal{M}}^{\text{R-indep.}}_\Delta(\ptv;\{k,l\}) \,
    \left(\frac{\nu}{k^-}\right)^\alpha \left(\frac{\nu}{l^- + l^+}\right)^\alpha
    + (k \leftrightarrow l)\,,
\end{align}
where we stress that the rapidity regulator is retained in its original form \eqref{eq:regulator} in the soft region. Due to the $n$-$\bar{n}$ symmetry of the process, there is no need to evaluate the mixing between anti-collinear and soft contributions explicitly, and for the collinear/anti-collinear mixing one finds a scaleless integral that vanishes in the applied regularisation prescription. Finally, for the double soft emissions one has
\begin{align}
\Delta {\cal S}_{ii'}^{(2)}(\ptv,\mu,\nu)  &= 
    \int \!\frac{\mathrm{d}^4 k}{(2\pi)^3} \, \frac{\mathrm{d}^4 l}{(2\pi)^3}  \; 
    \delta(k^2)\theta(k^0) \, \delta(l^2)\theta(l^0) \,
    \left(\frac{\nu}{k^-+k^+}\right)^\alpha \left(\frac{\nu}{l^- + l^+}\right)^\alpha
     \notag \\[0.5em]
    &\quad \times \mathcal{A}_{s}^{(2)}(k,l) \; 
    \widehat{\mathcal{M}}^{\text{R-indep.}}_\Delta(\ptv;\{k,l\}) \,,
\end{align}
where $\mathcal{A}_{s}^{(2)}$ denotes the soft double-emission amplitude, whose explicit expression can be found e.g.~in~\cite{Becher:2012qc}.

Focusing on concreteness on the diagonal gluon channel, the various contributions can be put together and the final correction to the refactorised matching kernel in \eqref{eq:Iren} can be obtained from
\begin{align}
     \Delta I_{g\leftarrow g}^{(2)}(z) &= \Delta {\cal I}_{g\leftarrow g}^{(2)}(z,\ptv,\mu,\nu) 
    + \Delta {\cal I}_{g\leftarrow g}^{(2,cs)}(z,\ptv,\mu,\nu)
    \nonumber\\
    & \qquad  + \frac12\, \Delta {\cal S}_{gg}^{(2)}(\ptv,\mu,\nu)  \, \delta(1-z)
    - \Delta d_2^A(R) \, \ln\frac{\ptv}{Q} \, \delta(1-z)\,,
\end{align}
where the last term subtracts the contribution from the two-loop anomaly exponent, which we decompose as
$\Delta d_2^A(R) = \Delta d_2^{C_A^2}(R) \,C_A^2 + \Delta d_2^{n_f}(R) \,C_A T_F n_f$. Explicitly, we find  
\begin{align}
  \Delta d_2^{C_A^2}(R) &= 
  \left(-\frac{524}{9} + \frac{16 \pi^2}{3} + \frac{176}{3} \ln 2 \right) \ln R + \frac{3220}{27} - \frac{44 \pi^2}{9} - \frac{560 }{9} \ln 2 - \frac{176 }{3}  \ln^2 2 - 48 \zeta_3
  \notag \\ 
  &\quad
  + 17.9 R^2 - 1.28 R^4 - 0.0169 R^6 + 0.000402 R^8  - 2.62 \cdot 10^{-5} R^{10}\,, 
  \notag \\
  \Delta d_2^{n_f}(R) &= \left(\frac{184}{9} - \frac{64}{3} \ln 2\right) \ln R -\frac{1256}{27} + \frac{16 \pi^2}{9} + \frac{256}{9} \ln 2 + \frac{64}{3}  \ln^2 2  \notag \\
  &\quad 
  - 0.706 R^2 + 0.0142 R^4 - 0.00217 R^6 + 0.000135 R^8 - 1.19 \cdot 10^{-5} R^{10} \,,
  \label{eq:delta-d2}
\end{align}
which agrees with previous calculations~\cite{Becher:2013xia}\footnote{Note that~\cite{Becher:2013xia} presents the difference to transverse-momentum resummation, whereas in our notation $\Delta d_2^A(R) $ refers to the difference to the integrated spectrum defined in~\eqref{eq:def:reference}.}. The extraction is less involved for the off-diagonal channels, and in each case, we can write the final expression for the refactorised matching kernels in the form
\begin{align}\label{eq:deltaI:res}
    \Delta I_{i\leftarrow j}^{(2)}(z) = 
    -  \delta_{ij} \, \Delta d_2^{R_i}(R)  \left[\frac{1}{1-z}\right]_+   
    + \delta_{ij} \, \Delta\calC_{\delta}^{R_i}(R) \, \delta(1-z) 
    + H_{i\leftarrow j}(z,R),
\end{align}
where the coefficient of the plus-distribution is given by the anomaly exponent in the representation $R_i$ of the parton $i$, and the coefficient of the delta function has a similar expansion in terms of colour factors and $R$-dependence. Particularly, we obtain for the diagonal gluon channel
\begin{align}
  \label{eq:DistributionsGG}
  \Delta\calC_{\delta}^{C_A^2}(R)  &= \left(-\frac{1622}{27} + \frac{548}{9} \ln 2 + \frac{88}{3} \ln^2 2 + 8 \zeta_3\right) \ln R -19.0 
  \notag\\
  &\quad
  + 4.60 R^2 + 0.478 R^4 - 0.0535 R^6 - 0.00128 R^8 + 1.35 \cdot 10^{-6} R^{10}\,, 
  \notag \\
  \Delta\calC_{\delta}^{n_f}(R)  &= \left(\frac{652}{27} - \frac{232}{9} \ln 2 - \frac{32}{3} \ln^2 2\right) \ln R + 1.27 
  \notag\\
  &\quad
  + 0.0538 R^2 - 0.0209 R^4 + 0.000599 R^6 - 0.000146 R^8 + 9.93 \cdot 10^{-6} R^{10}\,,
\end{align}
and the corresponding expressions for the diagonal quark channel can be found in App.~\ref{app:quark-bf}. Finally, we obtain purely numerical results for the non-distributional terms $H_{i\leftarrow j}(z,R)$ in \eqref{eq:deltaI:res} that depend on both the momentum fraction $z$ and the jet radius $R$.

We conclude this section by noting that our semi-analytical approach is similar in spirit to the method that was used in~\cite{Abreu:2022zgo}, but it differs in several technical aspects. First and foremost, our calculation uses a different rapidity regulator, which is purely analytic, and therefore all zero-bin subtractions are scaleless and vanish in our setup. Moreover, we use slightly different techniques for computing the uncorrelated-emission contribution, as well as a different reference observable than~\cite{Abreu:2022zgo} as described above.

%%%%%%%%%%%%%%%%%%%%%%%%%%%%%%%%%%%%%%%%%%%
\section{Results}
%%%%%%%%%%%%%%%%%%%%%%%%%%%%%%%%%%%%%%%%%%%
\label{sec:results}

In this section, we present our results for the non-logarithmic contributions to the renormalised matching kernels $I_{i\leftarrow j}^{(m)}(z)$ that are defined in \eqref{eq:Iren}. Within our novel approach, we can directly compute these kernels in momentum space, but for completeness, we also present the respective Mellin-space results that were obtained with the method described in~\cite{Bell:2022nrj}. While we focus here on the gluon channels with $i=g$, our novel momentum-space results for the quark kernels ($i=q$) are collected in App.~\ref{app:quark-bf}. As QCD is invariant under charge conjugation, the anti-quark kernels ($i=\bar q$) directly follow from these expressions.

%%%%%%%%%%%%%%%%%%%%%%%%%%%%%%%%%%%%
\subsection{Momentum-space kernels}
%%%%%%%%%%%%%%%%%%%%%%%%%%%%%%%%%%%%

At NLO, the algorithmic nature of the jet veto does not show up yet, and the matching kernels can be obtained analytically. Specifically, they read~\cite{Becher:2012qa}
\begin{align}
I_{g \leftarrow g}^{(1)}(z) &= 
C_A \,\bigg\{
-\frac{\pi^2}{6} \,\delta (1-z) 
\bigg\}
\,,
\nonumber\\
I_{g \leftarrow q}^{(1)}(z) &= 
C_F \,\bigg\{ 
2 z 
\bigg\}\,,
\end{align}
whereas the one-loop anomaly coefficient $d_1^A$ vanishes at this order.

%%%%%%%%%%%%%%%%%%%%%%%%%%%%%%%%%%%%%%%%%%%%%%%%%%%%%%%%%%%%%%%%%%%%%%%%
\begin{figure}[t!]
\centerline{
\includegraphics[height=5.2cm]{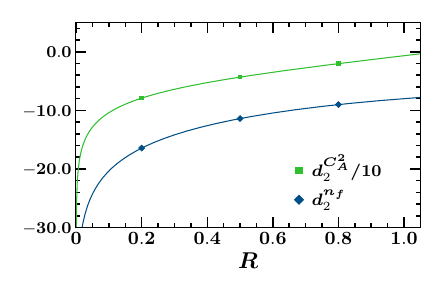}
\includegraphics[height=5.2cm]{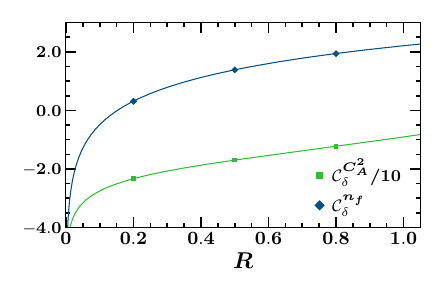}}
\vspace{-2mm}
\caption{\small{Left: Two-loop anomaly exponent $d_2^A(R) =  d_2^{C_A^2}(R) \,C_A^2 + d_2^{n_f}(R) \,C_A T_F n_f$ as a function of the jet radius $R$. The dots show the result of our numerical approach, and the lines represent the expression in~\eqref{eq:d2:final}. Right: The same for the coefficient of the delta function $\calC_{\delta}^A(R) =  \calC_{\delta}^{C_A^2}(R) \,C_A^2 + \calC_{\delta}^{n_f}(R) \,C_A T_F n_f$, where the lines refer to~\eqref{eq:Cdelta:final}.}}
\label{fig:DistributionGG}
\end{figure}
%%%%%%%%%%%%%%%%%%%%%%%%%%%%%%%%%%%%%%%%%%%%%%%%%%%%%%%%%%%%%%%%%%%%%%%%

At NNLO, we first verify if our setup reproduces the known results for the two-loop anomaly coefficient $d_2^A$ and the non-cusp anomalous dimension $\gamma_1^g$. As the former depends on the jet radius, we evaluate it numerically for three different values of $R\in\{0.2,0.5,0.8\}$ using the method described in Sec.~\ref{sec:numerical}. In the left panel of Fig.~\ref{fig:DistributionGG}, we compare these numbers that are indicated by the dots against the semi-analytical expression
\begin{align}
\label{eq:d2:final}
d_2^A(R)=&
C_A^2\left(\frac{808}{27}+4\zeta_3\right)
+C_A T_F n_f \left(- \frac{224}{27}\right)
+ C_A^2 \, \Delta d_2^{C_A^2}(R)
+ C_A T_F n_f \, \Delta d_2^{n_f}(R)\,,
\end{align}
where the first two terms originate from the reference observable \eqref{eq:def:reference}, and the latter two terms were given in \eqref{eq:delta-d2} above. The plot shows that the two calculations are in excellent agreement within the uncertainties that are too small to be visible on the scale of the plot. We also remark that the anomaly exponent exhibits Casimir scaling, and it is therefore related to the anomaly coefficient in the fundamental representation $d_2^F(R)$ that is relevant for Drell-Yan production. On the other hand, the non-cusp anomalous dimension is independent of the jet radius $R$ and it is constrained by RG consistency. Writing $\gamma_1^g = \gamma_1^{C_A^2} \,C_A^2 + \gamma_1^{C_A T_F} \,C_A T_F n_f+ \gamma_1^{C_F T_F} \,C_F T_F n_f$, we find
\begin{alignat}{2}
\gamma_1^{C_A^2}&= -17.1956 (141)\, &&\qquad
[-17.1941]\,, 
\no\\
\gamma_1^{C_A T_F}&= 7.2870 (6)\, &&\qquad
[7.2882]\,, 
\no\\
\gamma_1^{C_F T_F}&= 4.0000 (1)\, &&\qquad
[4]\,, 
\end{alignat}
where the numbers in square brackets show the known analytic results. The agreement provides another strong check of our computation.

Coming to the renormalised matching kernels $I_{i\leftarrow j}^{(2)}(z)$, we first consider their distribu\-tion-valued component. To this end, we decompose the kernels in the form
\begin{align}
    I_{i\leftarrow j}^{(2)}(z) = 
    -  \delta_{ij} \, d_2^{R_i}(R)  \left[\frac{1}{1-z}\right]_+   
    + \delta_{ij} \, \calC_{\delta}^{R_i}(R) \, \delta(1-z) 
    +  {I}_{i\leftarrow j}^{(2,\grid)}(z,R)\,,
    \label{eq:I2:distributions}
\end{align}
where the coefficient of the plus-distribution is given by the two-loop anomaly exponent in the representation $R_i$ of the parton $i$, and it thus carries no new information. The coefficient of the delta function, on the other hand, is for the gluon channel given by
\begin{align}
\calC_{\delta}^{A}(R)&=
C_A^2\left(\frac{1214}{81}-\frac{67\pi^2}{36}+\frac{5\pi^4}{72}+\frac{11}{9}\zeta_3\right)
+C_A T_F n_f \left(-\frac{328}{81}+\frac{5\pi^2}{9}-\frac{4}{9}\zeta_3\right)
\nonumber\\
&\quad + C_A^2 \; \Delta \calC_{\delta}^{C_A^2}(R)
+ C_A T_F n_f\; \Delta \calC_{\delta}^{n_f}(R) \,,
\label{eq:Cdelta:final}
\end{align}
with explicit expressions for the last two terms given in \eqref{eq:DistributionsGG}. Our numerical results for these coefficients are compared to these semi-analytical expressions in the right panel of Fig.~\ref{fig:DistributionGG}, which again shows perfect agreement between the two calculations.

%%%%%%%%%%%%%%%%%%%%%%%%%%%%%%%%%%%%%%%%%%%%%%%%%%%%%%%%%%%%%%%%%%%%%%%%
\begin{figure}[t!]
\centerline{
\includegraphics[height=5.2cm]{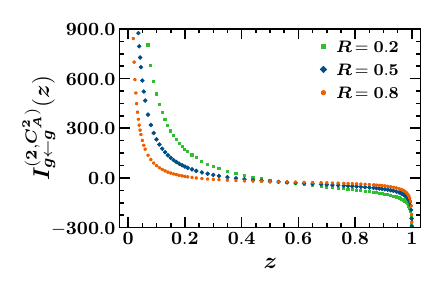}
\includegraphics[height=5.2cm]{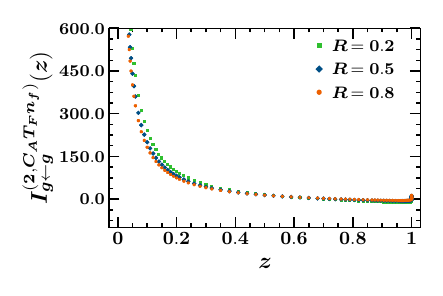}}
\vspace{-2mm}
\centerline{
\includegraphics[height=5.2cm]{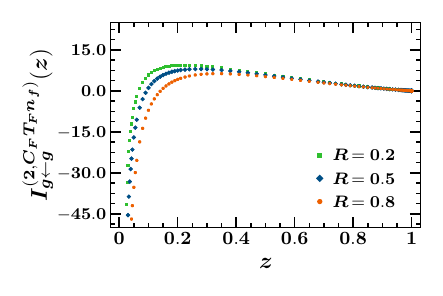}
\includegraphics[height=5.2cm]{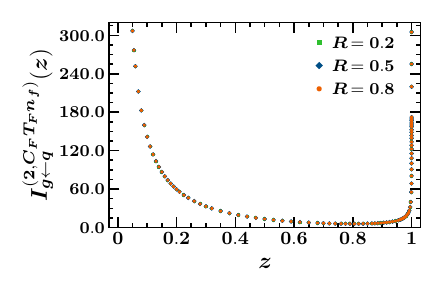}}
\vspace{-2mm}
\centerline{
\includegraphics[height=5.2cm]{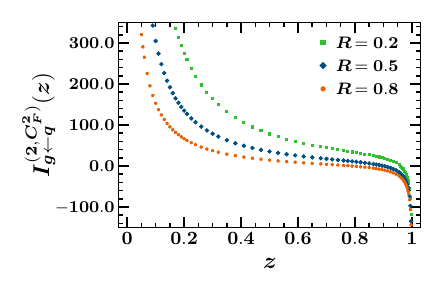}
\includegraphics[height=5.2cm]{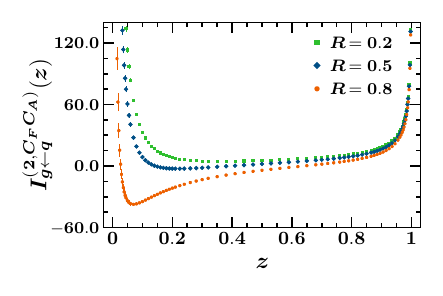}}
\vspace{-2mm}
\caption{\small{Grid contributions to the NNLO matching kernels  defined in \eqref{eq:I2:distributions} and \eqref{eq:NNLOkernels} for three different values of the jet radius $R$.}}
\label{fig:GluonNNLOkernels}
\end{figure}
%%%%%%%%%%%%%%%%%%%%%%%%%%%%%%%%%%%%%%%%%%%%%%%%%%%%%%%%%%%%%%%%%%%%%%%%

We finally turn to the grid contributions, which we decompose further in terms of their colour structure,
\begin{align}
{I}_{g\leftarrow g}^{(2,\grid)}(z,R) &=
C_A^2 \; {I}_{g\leftarrow g}^{(2,C_A^2)}(z) 
+ C_A T_F n_f\; {I}_{g\leftarrow g}^{(2,C_A T_F n_f)}(z) 
+ C_F T_F n_f \; {I}_{g\leftarrow g}^{(2,C_F T_F n_f)}(z) \,,
\nonumber\\
{I}_{g\leftarrow q}^{(2,\grid)}(z,R) &=
C_F T_F n_f \; {I}_{g\leftarrow q}^{(2,C_F T_F n_f)}(z) 
+ C_F^2 \; {I}_{g\leftarrow q}^{(2,C_F^2)}(z) 
+ C_F C_A \; {I}_{g\leftarrow q}^{(2,C_F C_A)}(z) \,,
\label{eq:NNLOkernels}
\end{align}
where we kept the $R$-dependence of the individual kernels on the right-hand-side of this equation implicit. We evaluated these kernels numerically for three different values of the jet radius $R\in\{0.2,0.5,0.8\}$ and 125 values of the momentum fraction $z$ using the method described in Sec.~\ref{sec:numerical}. The results are displayed in Fig.~\ref{fig:GluonNNLOkernels}, and they are also provided in electronic form in the file that accompanies the present article. Specifically, one notices that the off-diagonal kernel ${I}_{g\leftarrow q}^{(2,C_F T_F n_f)}(z)$ does not show any dependence on the jet radius, since it only receives real-virtual contributions, whereas all the other kernels exhibit a dependence on $R$, which is more pronounced for small values of the momentum fraction $z$. Similar to the plots in Fig.~\ref{fig:DistributionGG}, the numerical uncertainties are not visible on the scale of the plots, with the only exception being the low $z$-region of the off-diagonal ${I}_{g\leftarrow q}^{(2,C_F C_A)}(z)$ kernel. While these uncertainties are generically at the sub-percent level,  we observe accidental cancellations in this case, which make them somewhat more pronounced for this particular kernel.

%%%%%%%%%%%%%%%%%%%%%%%%%%%%%%%%%%%%%%%%%%%%%%%%%%%%%%%%%%%%%%%%%%%%%%%%
\begin{figure}[t!]
\centerline{
\includegraphics[height=5.6cm]{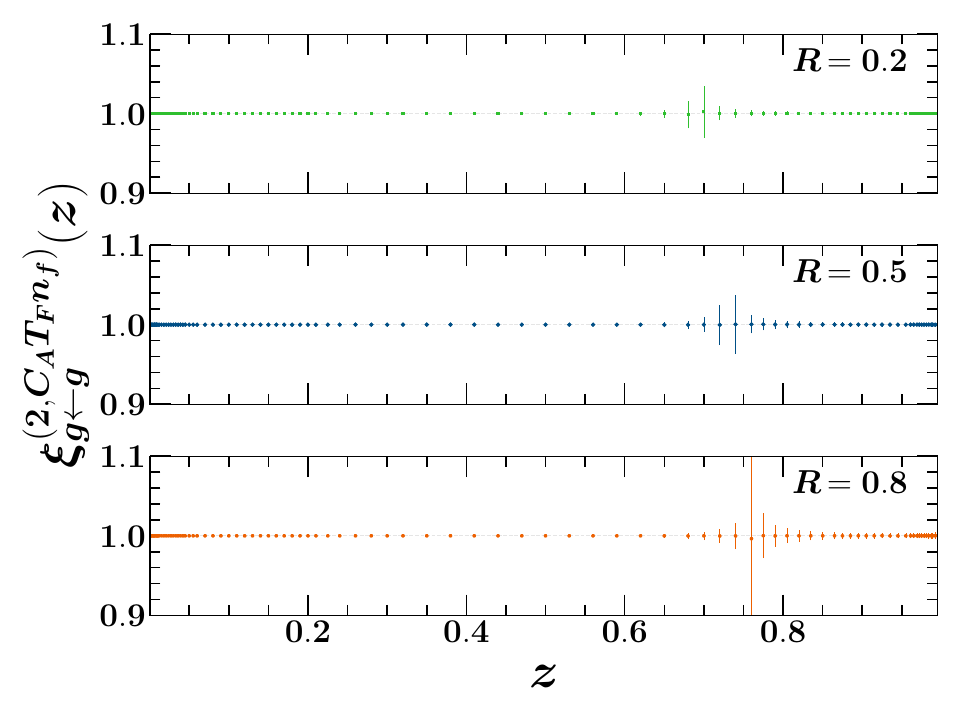}
\includegraphics[height=5.6cm]{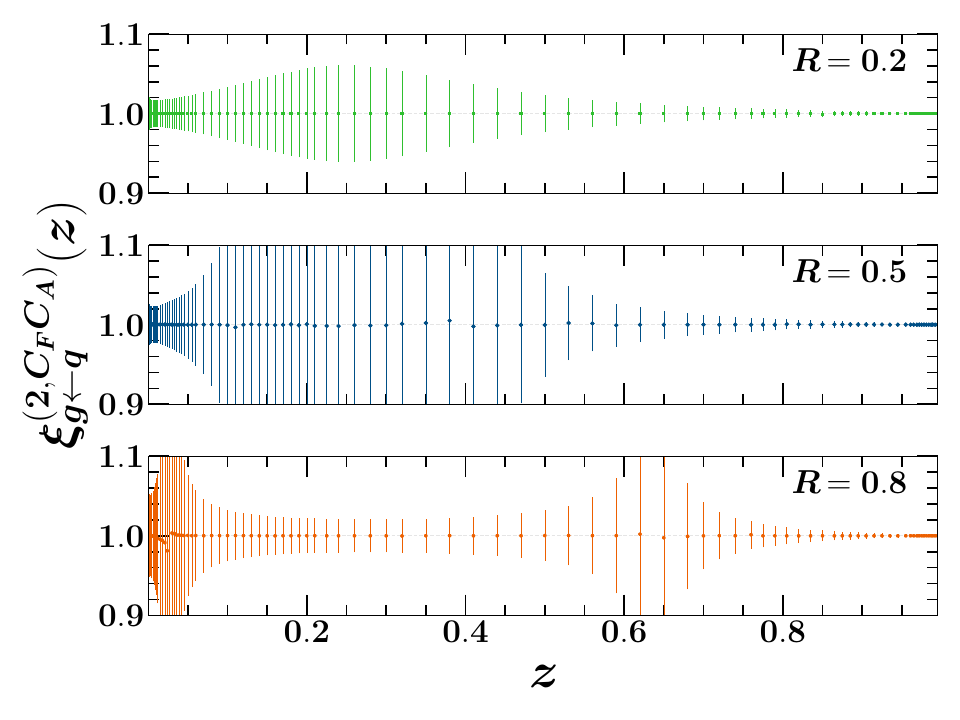}}
\vspace{1mm}
\centerline{
\includegraphics[height=5.6cm]{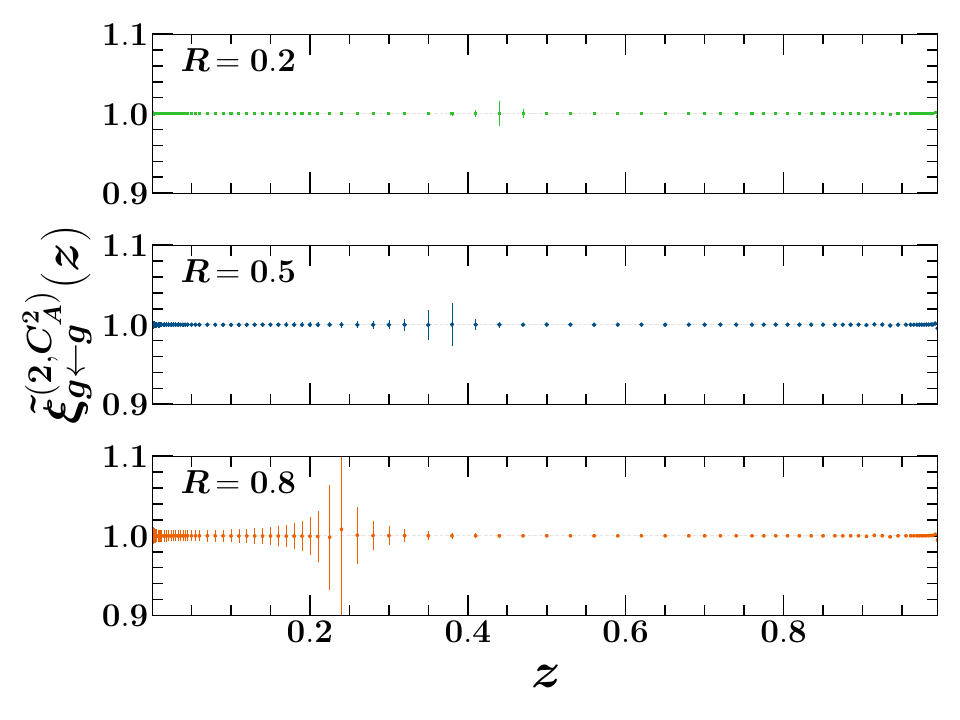}
\includegraphics[height=5.6cm]{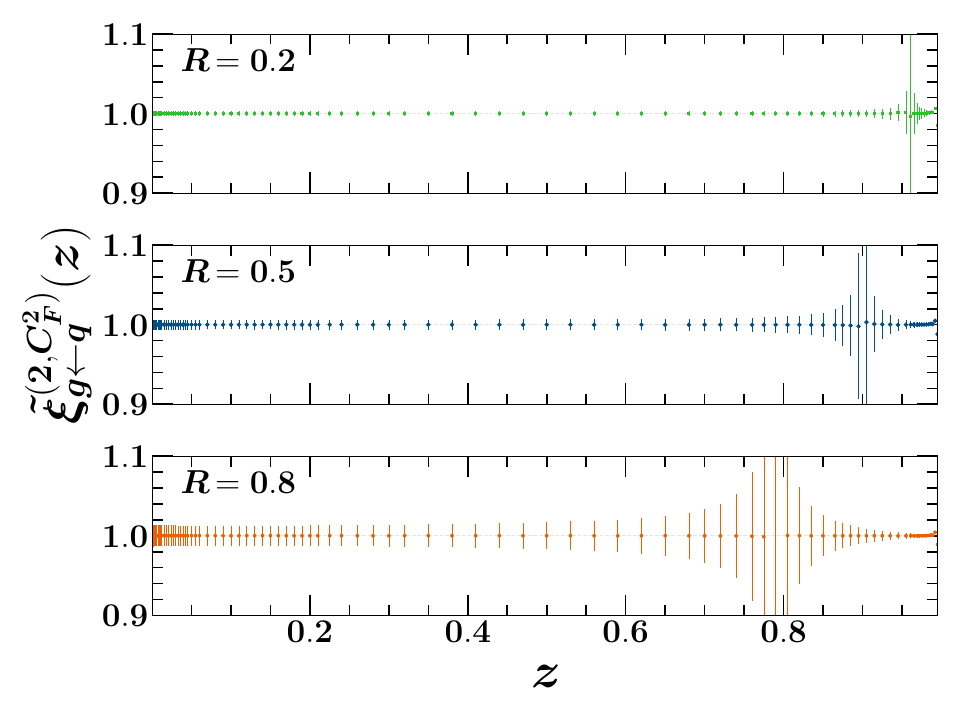}}
\vspace{-2mm}
\caption{\small{Upper: Ratio of the grid contributions between our numerical and our semi-analytical results as defined in \eqref{eq:ratioNA}. We show two sample matching kernels for three values of the jet \mbox{radius $R$}. Lower: The same for the ratio of the grid contributions between our numerical approach and the results of~\cite{Abreu:2022zgo} as defined in \eqref{eq:ratioNO}.}}
\label{fig:RatioNNLOkernels}
\end{figure}
%%%%%%%%%%%%%%%%%%%%%%%%%%%%%%%%%%%%%%%%%%%%%%%%%%%%%%%%%%%%%%%%%%%%%%%%

We next compare these numbers against the output of our semi-analytical approach. For this purpose, we define the following ratios involving the grid contributions of the two approaches,
\begin{align}
	\xi_{i\leftarrow j}^{(2,X)}(z)
	= \frac{\Big[{I}_{i\leftarrow j}^{(2,X)}(z)\Big]_{\text{Numerical~~~~~~~}}}{\Big[{I}_{i\leftarrow j}^{(2,X)}(z)\Big]_{\text{Semi-analytical}}}\,,
	\label{eq:ratioNA}
\end{align}
where the subscripts `Numerical' and `Semi-analytical' refer to the two setups of our calculation described in Sec.~\ref{sec:numerical} and Sec.~\ref{sec:analytical}, respectively, and the index $X$ labels one of the colour structures in the notation of~\eqref{eq:NNLOkernels}. We present a template comparison for one diagonal and one off-diagonal kernel in the upper panels of Fig.~\ref{fig:RatioNNLOkernels}. In general, we find a very good agreement between the two calculations, which supports our previous claim that the uncertainties of our numerical approach are at the sub-percent level. This can clearly be seen in the bulk of the distributions in the first panel of Fig.~\ref{fig:RatioNNLOkernels}. In these plots one also notices that the \emph{relative} uncertainties are enhanced in those regions where the central values of the kernels are close to zero. As argued above, the situation is special for the off-diagonal channel that is shown in the right panel of Fig.~\ref{fig:RatioNNLOkernels} because of accidental cancellations, and in this case one finds relative uncertainties of a few percent even in regions where the central values of the kernels are not small. We remark, however, that these uncertainties are likely to be overestimated, since the shifts of the central values are in all plots much smaller than what is indicated by the error estimate.

%%%%%%%%%%%%%%%%%%%%%%%%%%%%%%%%%%%%%%%%%%%%%%%%%%%%%%%%%%%%%%%%%%%%%%%%
\begin{figure}[t!]
\centerline{
\includegraphics[height=5.6cm]{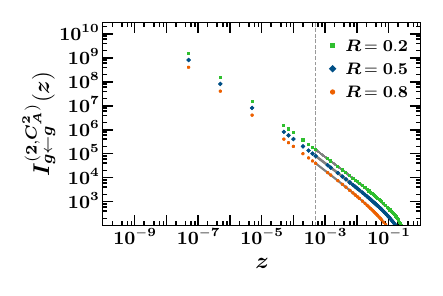}
\includegraphics[height=5.6cm]{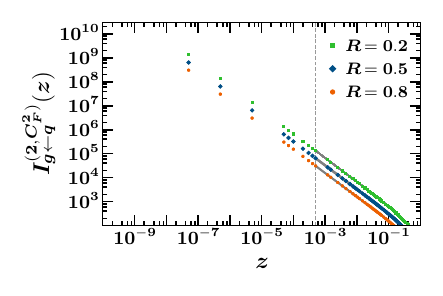}}
\vspace{-2mm}
\centerline{
\hspace{-2mm}
\includegraphics[height=5.7cm]{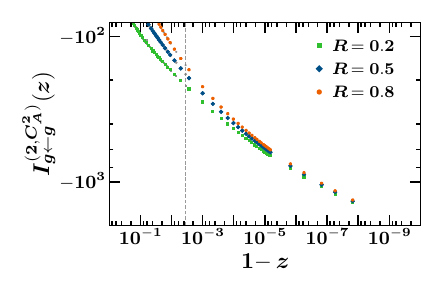}
\hspace{-3mm}
\includegraphics[height=5.7cm]{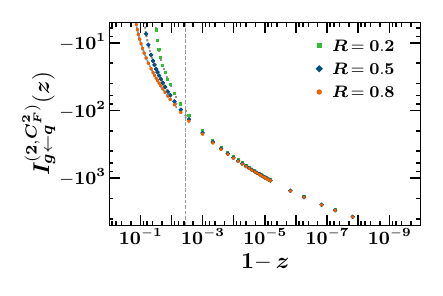}}
\vspace{-2mm}
\caption{\small{Endpoint regions $z\to 0$ (upper row) and $z\to 1$ (lower row) for the same matching kernels as in the lower panels of Fig.~\ref{fig:RatioNNLOkernels}, where our results are shown in colour and the ones from~\cite{Abreu:2022zgo} in gray. The smallest $z$ values (upper row) and the largest $z$ values (lower row)  of the grids provided in~\cite{Abreu:2022zgo} are indicated by the dashed vertical lines.}}
\label{fig:RatioNNLOkernelsNO}
\end{figure}
%%%%%%%%%%%%%%%%%%%%%%%%%%%%%%%%%%%%%%%%%%%%%%%%%%%%%%%%%%%%%%%%%%%%%%%%

As stated in the introduction, the jet-veto matching kernels were previously determined to NNLO in~\cite{Abreu:2022zgo}. Although our Mellin-space results for quark-initiated processes were already available at that time~\cite{Bell:2022nrj}, a comparison between the two calculations has not yet been presented. Here we perform such a comparison. Note that this requires to recombine the scheme-dependent kernels from~\cite{Abreu:2022zgo} with the corresponding soft function~\cite{Abreu:2022sdc} in order to extract the refactorised matching kernels according to \eqref{eq:refact}. Using \eqref{eq:Iren} and    \eqref{eq:I2:distributions} one then determines the grid contributions ${I}_{i\leftarrow j}^{(2,\grid)}(z,R)$, which we use for the comparison. Specifically, we define the ratios
\begin{align}
	\widetilde{\xi}_{i\leftarrow j}^{(2,X)}(z)
	= \frac{\Big[{I}_{i\leftarrow j}^{(2,X)}(z)\Big]_{\text{Numerical}}}{\Big[{I}_{i\leftarrow j}^{(2,X)}(z)\Big]_{\text{Ref.\ \cite{Abreu:2022zgo}~~}}}\,,
 	\label{eq:ratioNO}
\end{align}
which are displayed for two template kernels in the lower panels of Fig.~\ref{fig:RatioNNLOkernels}. As is evident from these plots, we find a very good agreement between the two calculations for all values of the momentum fraction $z$ (the pattern around the zeroes of the matching kernels being similar to the one in the upper panels). In order to verify if this agreement persists in the endpoint regions $z\to 0$ and $z\to 1$, where most of the kernels are enhanced, we show the same matching kernels in those regions in Fig.~\ref{fig:RatioNNLOkernelsNO}. While it becomes hard to distinguish our results (in colour) from the ones of~\cite{Abreu:2022zgo} (in gray) in these plots, we find  (i) that the agreement extends nicely into both endpoint regions, and (ii) that our numbers are stable even for extreme values of $z$, which allows one to sample these regions to very high accuracy.

%%%%%%%%%%%%%%%%%%%%%%%%%%%%%%%%%%%%
\subsection{Mellin-space kernels}
%%%%%%%%%%%%%%%%%%%%%%%%%%%%%%%%%%%%

%%%%%%%%%%%%%%%%%%%%%%%%%%%%%%%%%%%%%%%%%%%%%%%%%%%%%%%%%%%%%%%%%%%%%%%%
\begin{figure}[t!]
\centerline{
\includegraphics[height=5.2cm]{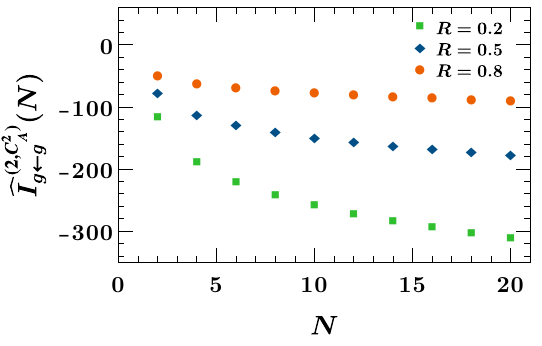}
\includegraphics[height=5.2cm]{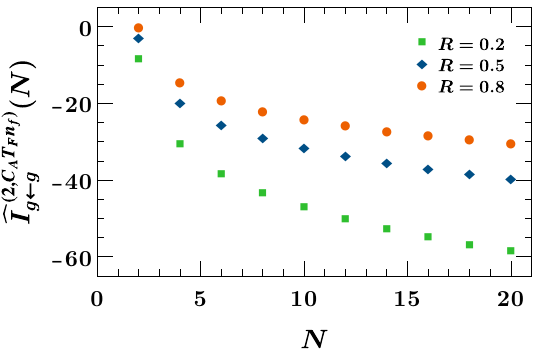}}
\vspace{1mm}
\centerline{
\hspace{2mm}
\includegraphics[height=5.2cm]{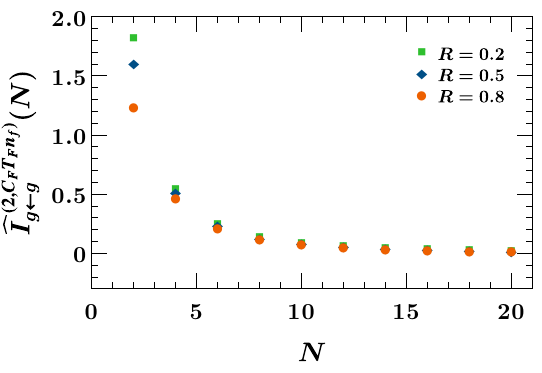}
\hspace{3mm}
\includegraphics[height=5.2cm]{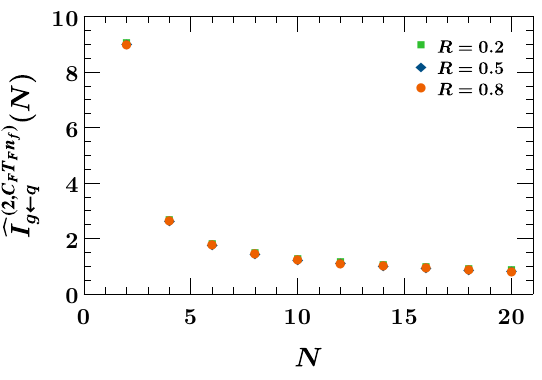}}
\vspace{1mm}
\centerline{
\includegraphics[height=5.2cm]{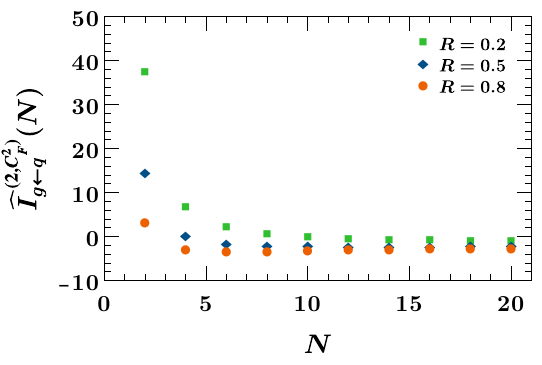}
\hspace{3mm}
\includegraphics[height=5.2cm]{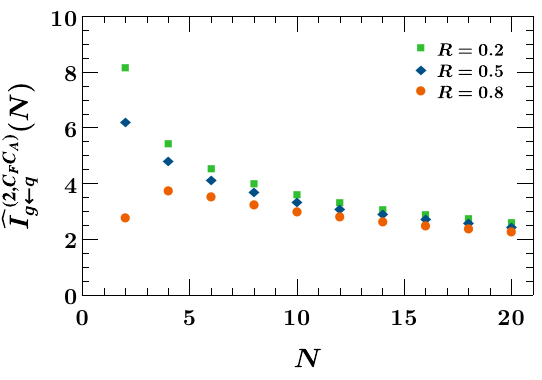}}
\vspace{-1mm}
\caption{\small{NNLO matching kernels for three different values of the jet radius $R$ in Mellin space.}}
\label{fig:GluonNNLOkernels:Mellin}
\end{figure}
%%%%%%%%%%%%%%%%%%%%%%%%%%%%%%%%%%%%%%%%%%%%%%%%%%%%%%%%%%%%%%%%%%%%%%%%

In a previous study~\cite{Bell:2022nrj}, a subset of the current authors computed the jet-veto matching kernels for quark-initiated processes at NNLO directly in Mellin space. These kernels are related to the refactorised matching kernels defined in \eqref{eq:refact} via
\begin{align}
   \widehat I_{i\leftarrow j}(N,\ptv,\mu) = \int_0^1 \df z \;
	z^{N-1} \;I_{i\leftarrow j}(z,\ptv,\mu)
	\,,
 \label{eq:MellinTransform}
\end{align}
and one similarly denotes the Mellin transform of the non-logarithmic contribution in \eqref{eq:I2:distributions} by $\widehat I_{i\leftarrow j}^{(2)}(N)$. While our results from the previous section can in principle be used for this conversion, the discretised grids would introduce systematic uncertainties in addition to the statistical errors from the Monte-Carlo integration. We therefore follow the strategy outlined in~\cite{Bell:2022nrj} here, and compute the Mellin-space kernels directly within our numerical approach by including an additional integration over the variable $z$ according to \eqref{eq:MellinTransform}. At NLO this integration can be performed analytically, and one obtains
\begin{align}
\widehat I_{g \leftarrow g}^{(1)}(N) &= 
C_A \,\bigg\{
-\frac{\pi^2}{6}
\bigg\}
\,,
\nonumber\\
\widehat I_{g \leftarrow q}^{(1)}(N) &= 
C_F \,\bigg\{ 
\frac{2}{N+1}
\bigg\}\,.
\end{align}
At NNLO, we then sample the Mellin-space kernels for ten values of the Mellin parameter $N\in\{2,4,6,8,10,12,14,16,18,20\}$ as in~\cite{Bell:2022nrj}. Writing the Mellin-space kernels in the form
\begin{align}
\widehat{I}_{g\leftarrow g}^{(2)}(N) &=
C_A^2 \; \widehat{I}_{g\leftarrow g}^{(2,C_A^2)}(N) 
+ C_A T_F n_f\; \widehat{I}_{g\leftarrow g}^{(2,C_A T_F n_f)}(N) 
+ C_F T_F n_f \; \widehat{I}_{g\leftarrow g}^{(2,C_F T_F n_f)}(N) \,,
\nonumber\\
\widehat{I}_{g\leftarrow q}^{(2)}(N) &=
C_F T_F n_f \; \widehat{I}_{g\leftarrow q}^{(2,C_F T_F n_f)}(N) 
+ C_F^2 \; \widehat{I}_{g\leftarrow q}^{(2,C_F^2)}(N) 
+ C_F C_A \; \widehat{I}_{g\leftarrow q}^{(2,C_F C_A)}(N) \,,
\end{align}
we show our results for the individual coefficients in this decomposition in Fig.~\ref{fig:GluonNNLOkernels:Mellin}. We note that the numerical uncertainties are again at the sub-percent level in this case, and they are therefore not visible on the scale of the plots. The Mellin-space kernels are particularly useful when the resummation is performed in Mellin space, as was done e.g.~for joint threshold and transverse-momentum resummation in~\cite{Kang:2022nft}.

%%%%%%%%%%%%%%%%%%%%%%%%%%%%%%%%%%%%%%%%%%%
\section{Conclusion}
%%%%%%%%%%%%%%%%%%%%%%%%%%%%%%%%%%%%%%%%%%%
\label{sec:conclusions}

We computed the beam-function matching kernels for jet-veto resummation to NNLO in QCD. Building on our earlier study of the respective quark beam function~\cite{Bell:2022nrj}, we provide the full set of matching kernels in this work. To this end, we applied two different computational methods, and we found perfect agreement between the two approaches. Our final results are displayed in Fig.~\ref{fig:GluonNNLOkernels} and Fig.~\ref{fig:QuarkNNLOkernels}, and they are also provided in electronic form as supplementary material to this paper. 

The jet-veto matching kernels were previously determined to NNLO in~\cite{Abreu:2022zgo}. We performed a detailed comparison to these results, paying special attention to the endpoint regions $z\to 0$ and $z\to 1$, and found very good agreement for all kernels. This represents the first confirmation of the results in~\cite{Abreu:2022zgo}. In addition, we for the first time computed the NNLO matching kernels for the gluon channels in Mellin space.

Whereas our semi-analytical method is specific to the jet-veto observable, our numerical approach is generic, and it can therefore be applied to a much broader class of observables. In contrast to our previous study~\cite{Bell:2022nrj}, in which we determined the matching kernels in Mellin space to avoid distribution-valued expressions, we have extended our formalism in this work such that it becomes capable of calculating the matching kernels directly in momentum space. We view this extension as a major improvement of our framework, and we anticipate many further applications in the future. In the long term, we plan to provide a public code for the computation of NNLO beam functions in the spirit of the {\tt SoftSERVE} distribution.

\acknowledgments
G.D.~thanks S.\ Jones for useful discussions related to {\tt pySecDec}. G.D.~also thanks the \texttt{ZIMT} support of the \texttt{OMNI} cluster of the University of Siegen where part of the computation was performed. Part of the simulation was performed by computing 
resources granted by RWTH Aachen University under the project  {\tt rwth1298}. The work of G.B., K.B., G.D.~and M.W. was supported by the Deutsche Forschungsgemeinschaft (DFG, German Research Foundation) under grant 396021762 - TRR 257 (\emph{``Particle Physics Phenomenology after the Higgs Discovery''}), while the work of D.Y.S.~was supported by the National Science Foundations of China under Grant No.~12275052 and No.~12147101, and the Shanghai Natural Science Foundation under Grant No.~21ZR1406100.

\begin{appendix}

%%%%%%%%%%%%%%%%%%%%%%%%%%%%%%%%%%%%%%%%%%%
\section{Anomalous dimensions and splitting functions}
%%%%%%%%%%%%%%%%%%%%%%%%%%%%%%%%%%%%%%%%%%%
\label{app:anoD}

We define the coefficients in the perturbative expansion of the anomalous dimensions appearing in the RG equations \eqref{eq:anomaly:RGE} and \eqref{eq:rgeI} via 
\begin{align}
\Gamma_{\mathrm{cusp}}^R(\alpha_s) = \sum_{m=0}^{\infty} \left(\frac{\alpha_{s}}{4 \pi} \right)^{m+1} \Gamma_{m}^R\,, \qquad 
\gamma^{i}(\alpha_s)= \sum_{m=0}^{\infty} \left(\frac{\alpha_{s}}{4 \pi} \right)^{m+1} \gamma^{i}_{m}\,,
\end{align} 
and likewise for the QCD $\beta$-function
\begin{align}
\beta(\alpha_s) = -2\alpha_s \sum_{m=0}^{\infty} \left(\frac{\alpha_{s}}{4 \pi} \right)^{m+1} \beta_{m}\,.
\end{align} 
The relevant coefficients of the cusp anomalous dimension are given by
\begin{align} 
\Gamma_0^i &= 4 C_i\,,
\nonumber \\
\Gamma_1^i &= 4 C_i
\left\{ \left(\frac{67}{9}-\frac{\pi^2}{3}\right) C_A
- \frac{20}{9} T_F n_f \right\}\,,
\end{align}
where $i=F$ refers to the fundamental and $i=A$ to the adjoint representation. Up to two-loop order, the collinear quark and gluon anomalous dimensions read~\cite{Becher:2006mr,Becher:2009qa},
\begin{align} 
\gamma_0^q &= -3C_F\,,
\nonumber\\
\gamma_1^q &=C_F^2 \bigg( -\frac32 + 2\pi^2 -24\zeta_3\bigg)
+ C_F C_A \bigg(-\frac{961}{54} - \frac{11\pi^2}{6} + 26\zeta_3\bigg) + C_F T_F n_f \bigg(\frac{130}{27}+\frac{2\pi^2}{3} \bigg),
\nonumber\\ 
\gamma_0^g &= -\frac{11}{3}C_A+\frac43 T_F n_f\,,
\nonumber\\ 
\gamma_1^g &= C_A^2 \bigg( -\frac{692}{27} + \frac{11\pi^2}{18} +2\zeta_3\bigg)
+ C_A T_F n_f \bigg(\frac{256}{27}-\frac{2\pi^2}{9} \bigg) + 4\,C_F T_F n_f \,,
\end{align}
whereas we only need the one-loop coefficient of the QCD $\beta$-function, $\beta_0 = \frac{11}{3} C_A - \frac43T_F n_f$, in this work.\\

\noindent
We furthermore expand the splitting functions entering \eqref{eq:rgeI} in the form
\begin{align}
P_{i\leftarrow j}(z,\als)=
\sum_{m=0}^{\infty} \left(\frac{\alpha_{s}}{4 \pi} \right)^{m+1} 
P_{i\leftarrow j}^{(m)}(z)\,.
\end{align} 
The relevant (non-vanishing) expressions at one-loop order are given by
\begin{align}
P_{q\leftarrow q}^{(0)}(z)&=
C_F \bigg\{\frac{4}{(1-z)_+}  + 3\, \delta(1-z) - 2 (1+z)\bigg\}\,,
\nonumber\\
P_{q\leftarrow g}^{(0)}(z)&=
2T_F \bigg\{ z^2 + (1-z)^2 \bigg\}\,,
\nonumber\\
P_{g\leftarrow q}^{(0)}(z)&=
2C_F \bigg\{ \frac{1+(1-z)^2}{z}  \bigg\}\,,
\nonumber\\
P_{g\leftarrow g}^{(0)}(z)&=
4C_A \bigg\{ \frac{1}{(1-z)_+} + \frac{1}{z}-2 + z(1-z)  \bigg\} + \beta_0\, \delta(1-z) \,,
\end{align} 
whereas the two-loop splitting functions can be found in~\cite{Curci:1980uw,Furmanski:1980cm,Ellis:1996nn}. Decomposing
\begin{align}
P_{q\leftarrow q}^{(1)}(z)&=
C_F^2 \; P_{q\leftarrow q}^{(1,C_F)}(z)
+ C_F C_A \; P_{q\leftarrow q}^{(1,C_A)}(z)   
+ C_F T_F n_f \; P_{q\leftarrow q}^{(1,n_f)}(z) 
+ C_F T_F \; P_{q\leftarrow q}^{(1,T_F)}(z)\,,
\nonumber\\[0.2em]
P_{q\leftarrow g}^{(1)}(z) &=
C_F T_F \; P_{q\leftarrow g}^{(1,C_F)}(z) 
+ C_A T_F \; P_{q\leftarrow g}^{(1,C_A)}(z) \,,
\nonumber\\[0.1em]
P_{q\leftarrow \bar q}^{(1)}(z) &=
C_F (C_A - 2 C_F) \; P_{q\leftarrow \bar q}^{(1,C_{AF})}(z) + C_F T_F \; P_{q\leftarrow q}^{(1,T_F)}(z) \,,
\nonumber\\[0.2em]
P_{q\leftarrow q'}^{(1)}(z) &=
P_{q\leftarrow \bar q'}^{(1)}(z) =
C_F T_F \; P_{q\leftarrow q}^{(1,T_F)}(z)\,,
\nonumber\\[0.1em]
P_{g\leftarrow q}^{(1)}(z) &=
C_F T_F n_f \; P_{g\leftarrow q}^{(1,C_F T_F n_f)}(z) 
+ C_F^2 \; P_{g\leftarrow q}^{(1,C_F^2)}(z) 
+ C_F C_A \; P_{g\leftarrow q}^{(1,C_F C_A)}(z) \,,
\nonumber\\[0.1em]
P_{g\leftarrow g}^{(1)}(z) &=
C_A^2 \; P_{g\leftarrow g}^{(1,C_A^2)}(z) 
+ C_A T_F n_f\; P_{g\leftarrow g}^{(1,C_A T_F n_f)}(z) 
+ C_F T_F n_f \; P_{g\leftarrow g}^{(1,C_F T_F n_f)}(z) \,,
\end{align}
in analogy to \eqref{eq:NNLOkernels} and \eqref{eq:gridQuark}, they read explicitly
\begin{align}
   P_{q\leftarrow q}^{(1,C_F)}(z) =& 
   \bigg( \frac32 - 2\pi^2 +24\zeta_3\bigg) \delta(1-z)
   -2 (1+z) \ln^2 z
   -\frac{8 (1+z^2)}{1-z} \ln z \,\ln (1-z) 
  \nonumber\\
   & - \frac{4(3 + 2 z - 2 z^2)}{1 - z} \ln z
   -20 (1-z)\,,
  \nonumber\\
  P_{q\leftarrow q}^{(1,C_A)}(z) =& 
    \bigg( \frac{268}{9} - \frac{4 \pi^2}{3}\bigg) \left[\frac{1}{1-z}\right]_+
     +\bigg(\frac{17}{6}+\frac{22 \pi ^2}{9}-12 \zeta_3\bigg) \delta (1-z)
    \nonumber\\
   & 
   +\frac{2 (1+z^2)}{1-z} \ln^2 z
     +\frac{2(17+5 z^2)}{3(1-z)} \ln z
     + \frac{2 \pi^2}{3} (1 + z) + \frac{106 - 374 z}{9}\,,
  \nonumber\\
  P_{q\leftarrow q}^{(1,n_f)}(z) =&
  -\frac{80}{9} \left[\frac{1}{1-z}\right]_+
  -\bigg( \frac23 + \frac{8 \pi^2}{9} \bigg) \delta (1-z)
  -\frac{8 (1+z^2)}{3 (1-z)} \ln z
  -\frac{8(1-11 z)}{9}\,,
  \nonumber\\
  P_{q\leftarrow q}^{(1,T_F)}(z) =&
   -4 (1+z) \ln^2 z
   +\frac{2 (6 + 30 z + 16 z^2)}{3} \ln z
   + \frac{8 (1 - z) (10 + z + 28 z^2)}{9z}\,,
   \nonumber\\
 P_{q\leftarrow g}^{(1,C_F)}(z) =&
 2(1-2z+2 z^2) \bigg(\!\ln^2 z -4 \ln z\, \ln (1-z)+2 \ln^2 (1-z) -\frac{2\pi^2}{3}\!\bigg)\!
 +4 z^2 \ln^2 z
  \nonumber\\
   & 
 +2(3-4 z + 8z^2) \ln z
 +16z (1-z) \ln (1-z)
 + 2(14 -29z +20 z^2)\,,
 \nonumber\\
 P_{q\leftarrow g}^{(1,C_A)}(z) =&
 -8 (1+2 z+2z^2) \Big(\text{Li}_2(-z)+\ln z\,\ln(1+z) \Big)
 -4(1+2 z) \ln^2 z
  \nonumber\\
   & 
   -4(1-2z+2z^2) \ln ^2(1-z)
    +\frac{4 (3 + 24 z + 44 z^2)}{3} \ln z
    -\frac{8\pi^2}{3}z
  \nonumber\\
   & 
    -16 z(1-z)  \ln (1-z)
    +\frac{4 (20 - 18 z + 225 z^2 - 218 z^3)}{9 z}\,,
  \nonumber\\
 P_{q\leftarrow \bar q}^{(1,C_{AF})}(z) =&
 \frac{2 (1+z^2)}{1+z} \bigg(4\text{Li}_2(-z)+4\ln z\,\ln(1+z) - \ln^2 z 
 +\frac{\pi^2}{3}\bigg)
 -4 (1+z) \ln z
 \nonumber\\
   & 
 -\frac{24}{3} (1 - z)\,,
 \nonumber\\
 P_{g\leftarrow q}^{(1,C_F T_F n_f)}(z)  =&
 -\frac{16 (2-2z+z^2)}{3 z} \ln (1-z)
 -\frac{32 (5-5z+4z^2)}{9 z}\,,
  \nonumber\\
 P_{g\leftarrow q}^{(1,C_F^2)}(z) =&
 -2 (2-z)\ln^2 z
 - \frac{4(2-2z+z^2)}{z} \ln^2 (1-z)
 +2(4+7 z) \ln z
 \nonumber\\
   & 
    -\frac{4 (6-6z+5 z^2)}{z} \ln (1-z)
    -2 (5+7 z)\,,
  \nonumber\\
P_{g\leftarrow q}^{(1,C_F C_A)}(z) =&
    ~\frac{8 (2+2z+z^2)}{z} \Big( \text{Li}_2(-z) + \ln z \, \ln (1+z) \Big)
    -\frac{4(36+15z+8z^2)}{3}  \ln z
 \nonumber\\
   & 
    -\frac{4(2-2z+z^2)}{z} \Big( 2\ln z \, \ln (1-z) - \ln^2 (1-z) \Big)
    + 4 (2+z)  \ln^2 z
\nonumber\\
   & 
    + \frac{4(22-22z+17 z^2)}{3z} \ln (1-z)
    +\frac{4 (9 + 19 z + 37 z^2 + 44 z^3)}{9 z}+\frac{8\pi^2}{3}\,,
  \nonumber\\
P_{g\leftarrow g}^{(1,C_A^2)}(z) =&
  \bigg( \frac{268}{9} - \frac{4 \pi^2}{3}\bigg) \left[\frac{1}{1-z}\right]_+ \!\!
  +\!\bigg(\frac{32}{3}+12 \zeta_3\bigg) \delta (1-z)
  -\frac{4(25-11 z+44 z^2)}{3} \ln z
  \nonumber\\
   & 
  +\frac{16 (1+z+z^2)^2}{z (1+z)}  \Big( \text{Li}_2(-z) + \ln z  \, \ln (1+z) \Big)
  + \frac{4 \pi^2 (3 + 4 z + 2 z^2 + 2 z^3)}{3 (1 + z)}
  \nonumber\\
   & 
    -\frac{16 (1-z+z^2)^2}{z(1-z)} \ln z \,  \ln (1-z)
  +\frac{8 (1+z-z^2)^2}{1-z^2} \ln^2 z
  -\frac{2(25 + 109 z)}{9}\,,
  \nonumber\\
P_{g\leftarrow g}^{(1,C_A T_F n_f)}(z) =&
-\frac{80}{9} \left[\frac{1}{1-z}\right]_+ 
  -\frac{16}{3} \, \delta (1-z)
  -\frac{16(1+z)}{3}  \ln z
  \nonumber\\
   & 
  -\frac{8 (23-29 z+19 z^2-23 z^3)}{9 z}
  \nonumber\\
P_{g\leftarrow g}^{(1,C_F T_F n_f)}(z) =&
   -4\, \delta (1-z)
   -8 (1+z) \ln ^2 z
   -8 (3+5 z) \ln z
   +\frac{16(1-12z+6 z^2+5 z^3)}{3z} \,.
\end{align}
The two-loop solutions of the RG equations given in \eqref{eq:Zfren} and \eqref{eq:Iren} involve convolutions of one-loop splitting functions. In the notation introduced in \eqref{eq:def:convolution}, which implies a sum over partons with index $l$, one has
\begin{align}
 &   \Big( P_{q\leftarrow l}^{(0)} \otimes  P_{l\leftarrow q}^{(0)} \Big) (z) 
\nonumber\\
&\quad  = C_F^2 \Bigg\{
32 \left[\frac{\ln(1-z)}{1-z}\right]_+
+24 \left[\frac{1}{1-z}\right]_+
+\bigg(9-\frac{8 \pi ^2}{3}\bigg) \delta(1-z)
- \frac{4(1+3 z^2)}{1-z} \ln z
\nonumber\\
&\quad\quad
-16 (1+z) \ln (1-z)
-4 (5+z)
\Bigg\} 
+ C_F T_F \Bigg\{
8 (1+z) \ln z
+\frac{4 (1-z) (4+7 z+4 z^2) }{3 z} 
\Bigg\} ,
\nonumber\\
&   \Big( P_{q\leftarrow l}^{(0)} \otimes  P_{l\leftarrow g}^{(0)} \Big) (z) 
\nonumber\\
&\quad  = 
C_F T_F \Bigg\{
8(1-2 z+2z^2 ) \ln (1-z)
-4(1-2 z+4z^2 ) \ln z
+8 z-2
\Bigg\} 
\nonumber\\
&\quad\quad  + C_A T_F \Bigg\{
8 (1-2 z+2 z^2) \ln (1-z)
+8(1+4 z) \ln z
+\frac{4(4+3z+24 z^2-31 z^3)}{3z} 
\Bigg\} 
\nonumber\\
&\quad\quad
+\beta_0 T_F\Bigg\{ 
2-4 z+4 z^2
\Bigg\} ,
\nonumber\\
&   \Big( P_{g\leftarrow l}^{(0)} \otimes  P_{l\leftarrow q}^{(0)} \Big) (z) 
\nonumber\\
&\quad  = 
C_A C_F \Bigg\{
 \frac{8(2-2z+z^2)}{z} \ln (1-z)
-\frac{16 (1+z+z^2)}{z}  \ln z
-\frac{4 (31 - 24 z - 3 z^2 - 4 z^3)}{3 z}
\Bigg\}
\nonumber\\
&\quad\quad + C_F^2 \Bigg\{
\frac{8 (2-2z+z^2)}{z} \ln (1-z)
+4(2- z) \ln z +8 -2 z
\Bigg\}\!
 + \beta_0 C_F \Bigg\{
\frac{2 (2-2z+z^2)}{z} 
\Bigg\},
\nonumber\\
&   \Big( P_{g\leftarrow l}^{(0)} \otimes  P_{l\leftarrow g}^{(0)} \Big) (z) 
\nonumber\\
&\quad  = 
C_A^2
   \Bigg\{
   32 \left[\frac{\ln(1-z)}{1-z}\right]_+
    -\frac{8 \pi ^2}{3}\, \delta(1-z)
    +\frac{32 (1-2z+z^2-z^3)}{z} \ln(1-z)
\nonumber\\
&\quad\quad
- \frac{16 (1 + 3 z^2 - 4 z^3 + z^4)}{z(1 - z)} \ln z
-\frac{16 (1 - z) (11 + 2 z + 11 z^2}{3z}
\Bigg\}
\nonumber\\
&\quad\quad
+ \beta_0 C_A \Bigg\{
8 \left[\frac{1}{1-z}\right]_+
+\frac{8 (1 - 2 z + z^2 - z^3)}{z}
\Bigg\}
+ \beta_0^2 \, \delta(1-z)
\nonumber\\
&\quad\quad
+ C_F T_F n_f \Bigg\{
8 (1+z) \ln z
+\frac{4 (1-z) (4+7 z+4 z^2) }{3 z} 
\Bigg\}\,,
\end{align}
where the factor $n_f$ in the last line arises from a sum over intermediate massless quarks. Similarly, the convolutions of the one-loop splitting functions with the one-loop jet-veto matching kernels read
\begin{align}
 &   \Big( I_{q\leftarrow l}^{(1)} \otimes  P_{l\leftarrow q}^{(0)} \Big) (z) 
\nonumber\\
&\quad  = 
-\frac{\pi^2}{6} C_F P_{q \leftarrow q}^{(0)}(z)
+C_F^2 \Bigg\{
2(1-z)\Big(
 4\ln (1-z)
-2 \ln z
-1
\Big) 
\Bigg\}
\nonumber\\
&\quad\quad
+ C_F T_F \Bigg\{
-8 z \ln z + \frac{8 (1 - z) (1 - 2 z - 2 z^2)}{3 z}
\Bigg\} ,
\nonumber\\
&   \Big( I_{q\leftarrow l}^{(1)} \otimes  P_{l\leftarrow g}^{(0)} \Big) (z) 
\nonumber\\
&\quad  = 
-\frac{\pi^2}{6} C_F P_{q \leftarrow g}^{(0)}(z)
+ C_F T_F \Bigg\{
-4 (1+2 z) \ln z - 4 (2-z-z^2)
\Bigg\} 
+ \beta_0 T_F \Bigg\{4z (1 - z) \Bigg\} 
\nonumber\\
&\quad\quad
  + C_A T_F \Bigg\{
 16 z (1 - z) \ln (1-z) 
 - 32 z \ln z 
 + \frac{8(1 - 3z- 15 z^2 + 17 z^3)}{3z}
 \Bigg\} ,
\nonumber\\
&   \Big( I_{g\leftarrow l}^{(1)} \otimes  P_{l\leftarrow q}^{(0)} \Big) (z) 
\nonumber\\
&\quad  = 
-\frac{\pi^2}{6} C_A P_{g \leftarrow q}^{(0)}(z)
 + C_F^2 \Bigg\{
 8 z \ln (1-z)-4 z \ln z+2z+4
\Bigg\},
\nonumber\\
&   \Big( I_{g\leftarrow l}^{(1)} \otimes  P_{l\leftarrow g}^{(0)} \Big) (z) 
\nonumber\\
&\quad  = 
-\frac{\pi^2}{6} C_A P_{g \leftarrow g}^{(0)}(z)
+ C_F T_F n_f \Bigg\{
8 z \ln z +4+4z-8 z^2
\Bigg\}\,,
\end{align}
where the factor $n_f$ in the last term again arises from a sum over all massless quark flavours.

%%%%%%%%%%%%%%%%%%%%%%%%%%%%%%%%%%%%%%%%%%%
\section{Details on the reference observable}
%%%%%%%%%%%%%%%%%%%%%%%%%%%%%%%%%%%%%%%%%%%
\label{app:reference}

Within the approach described in Sec.~\ref{sec:analytical}, the jet-veto matching kernels are computed starting from the difference \eqref{eq:deltaI} with respect to a reference observable, for which we use the integrated $q_T$-spectrum according to \eqref{eq:def:reference}. One then needs to add the matching kernels for this reference observable to the final expressions for the refactorised matching kernels given in \eqref{eq:deltaI:res}.

The matching kernels for the reference observable can easily be derived from available results for transverse-momentum resummation. As the latter are usually given in position-space, one needs to Fourier-transform these expressions and to integrate them up to the $\ptv$ cut. Specifically, logarithms in the transverse-momentum framework then translate into $\ptv$ logarithms using the relation
\begin{align}
& \int_0^{\ptv} \!\!\!\!\df q_T \; 2q_T\; \frac{1}{4\pi}
\int_0^{\infty} \df x_T \; x_T
\int_0^{2\pi} \df \varphi\;\;
e^{i q_T x_T \cos \varphi}\;
\ln^n \Big( \frac{x_T^2\mu^2}{4e^{-2\gamma_E}}\Big)
\nonumber\\
& \qquad
= \ptv 
\int_0^{\infty} \df x_T \; J_1\big(x_T \,\ptv\big) \;
\ln^n \Big( \frac{x_T^2\mu^2}{4e^{-2\gamma_E}}\Big)\,,
\end{align}
where $J_1(x)$ is a Bessel function. For $n=0,1,2$ there is thus a one-to-one correspondence between the two types of logarithms, whereas for $n=3,4$ they differ by certain $\zeta_3$ terms. This implies that the refactorised matching kernels for the reference observable are not precisely equal to the ones from transverse-momentum resummation for the diagonal channels. Particularly, we find 
\begin{align}
I_{i\leftarrow j}^{\text{ref},(2)}(z) =     
I_{i\leftarrow j}^{q_T,(2)}(z)
- \delta_{ij} \,\zeta_3 \, \Gamma_0^i \bigg\{ 2 P_{i\leftarrow j}^{(0)}(z) 
+\Big(d_1^i + 2 \gamma_0^i -\frac23 \beta_0\Big) \,\delta(1-z) \bigg\}\,,
\end{align}
where $I_{i\leftarrow j}^{q_T,(2)}(z)$ are the matching kernels for transverse-momentum resummation, for which we use the explicit expressions provided in~\cite{Gehrmann:2014yya} (setting the RG logarithms therein to zero).

%%%%%%%%%%%%%%%%%%%%%%%%%%%%%%%%%%%%%%%%%%%
\section{Quark beam function in momentum space}
%%%%%%%%%%%%%%%%%%%%%%%%%%%%%%%%%%%%%%%%%%%
\label{app:quark-bf}

%%%%%%%%%%%%%%%%%%%%%%%%%%%%%%%%%%%%%%%%%%%%%%%%%%%%%%%%%%%%%%%%%%%%%%%%
\begin{figure}[t!]
\centerline{
\includegraphics[height=5.2cm]{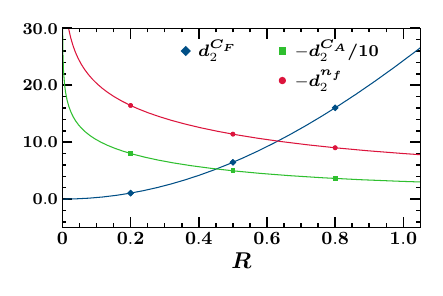}
\includegraphics[height=5.2cm]{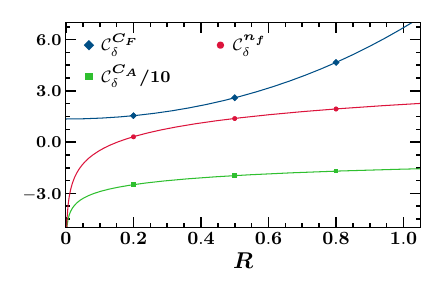}}
\vspace{-4mm}
\caption{\small{Left: Two-loop anomaly exponent $d_2^F = d_2^{C_F} \,C_F^2 + d_2^{C_A} \,C_F C_A + d_2^{n_f} \,C_F T_F n_f$ as a function of the jet radius $R$. The dots show the result of our numerical approach, and the lines represent the expression in~\eqref{eq:d2:QQ:final}. Right: The same for the coefficient of the delta function $\calC_{\delta}^F(R) = \calC_{\delta}^{C_F} \,C_F^2 + \calC_{\delta}^{C_A} \,C_F C_A + \calC_{\delta}^{n_f} \,C_F T_F n_f$, where the lines refer to~\eqref{eq:Cdelta:QQ:final}.}}
\label{fig:DistributionQQ}
\end{figure}
%%%%%%%%%%%%%%%%%%%%%%%%%%%%%%%%%%%%%%%%%%%%%%%%%%%%%%%%%%%%%%%%%%%%%%%%

In~\cite{Bell:2022nrj} we presented the first computation of the quark beam function for jet-veto resummation in Mellin space. Here we complement these results by calculating the matching kernels directly in momentum space using the two approaches described in Sec.~\ref{sec:computation}.

%%%%%%%%%%%%%%%%%%%%%%%%%%%%%%%%%%%%%%%%%%%%%%%%%%%%%%%%%%%%%%%%%%%%%%%%
\begin{figure}[t!]
\centerline{
\includegraphics[height=5.2cm]{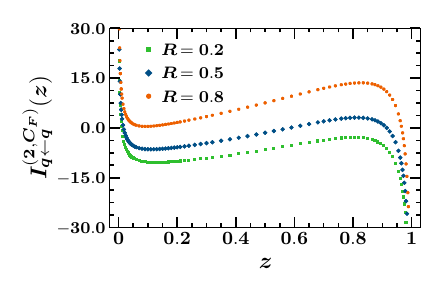}
\includegraphics[height=5.2cm]{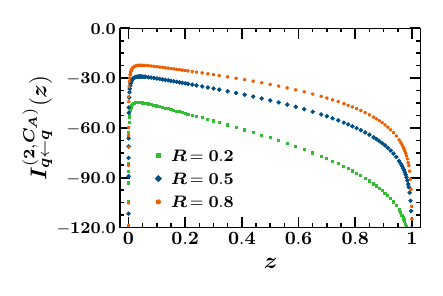}}
\vspace{-2mm}
\centerline{
\includegraphics[height=5.2cm]{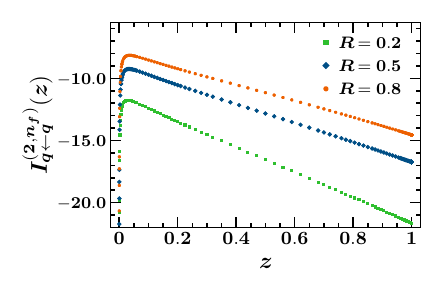}
\includegraphics[height=5.2cm]{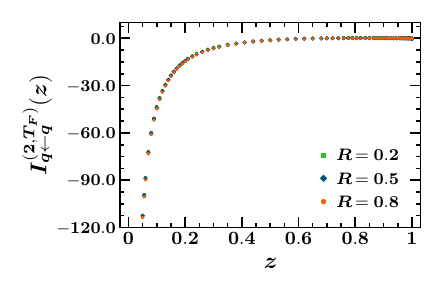}}
\vspace{-2mm}
\centerline{
\includegraphics[height=5.2cm]{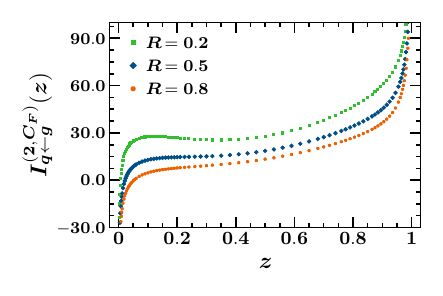}
\includegraphics[height=5.2cm]{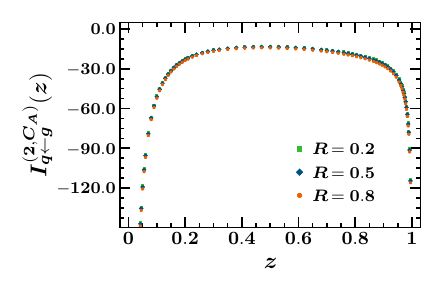}}
\vspace{-2mm}
\centerline{
\includegraphics[height=5.2cm]{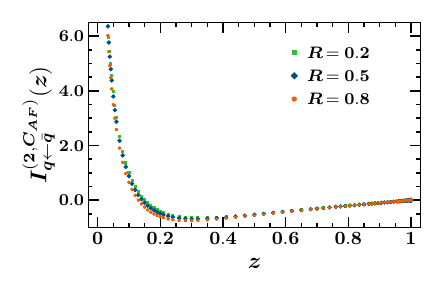}}
\vspace{-2mm}
\caption{\small{Grid contributions to the NNLO matching kernels  defined in \eqref{eq:gridQuark} for three different values of the jet radius $R$.}}
\label{fig:QuarkNNLOkernels}
\end{figure}
%%%%%%%%%%%%%%%%%%%%%%%%%%%%%%%%%%%%%%%%%%%%%%%%%%%%%%%%%%%%%%%%%%%%%%%%

At NLO the matching kernels can be obtained analytically. They read~\cite{Shao:2013uba}
\begin{align}
I_{q \leftarrow q}^{(1)}(z) &= 
C_F \,\left\{
-\frac{\pi^2}{6}\delta (1-z) + 2(1-z)
\right\}\,,
\nonumber\\
I_{q \leftarrow g}^{(1)}(z) &= 
T_F \bigg\{
4 z(1-z)
\bigg\}\,,
\end{align}
whereas we decompose the NNLO kernels in terms of distributions and a grid contribution according to~\eqref{eq:I2:distributions}. Our results for the coefficients of the distributions are shown in Fig.~\ref{fig:DistributionQQ}, where the numbers indicated by the dots have been obtained within our numerical approach from Sec.~\ref{sec:numerical}, and the lines correspond to
\begin{align}
d_2^F(R)&=
 C_F C_A\bigg(\frac{808}{27}-28\zeta_3\bigg)
+C_F T_F n_f \bigg(- \frac{224}{27}\bigg)
+ C_F^2\bigg(32\zeta_3\bigg)
\nonumber\\[0.0em]
&\quad
+ C_F^2 \, \Delta d_2^{C_F}(R)
+ C_F C_A \, \Delta d_2^{C_A}(R)
+ C_F T_F n_f \, \Delta d_2^{n_f}(R)\,,
\label{eq:d2:QQ:final}
\\[0.2em]
\calC_{\delta}^{F}(R)&=
 C_F C_A\left(\frac{1214}{81}-\frac{67\pi^2}{36}+\frac{\pi^4}{18}+\frac{11}{9}\zeta_3\right)
+C_F T_F n_f \left(-\frac{328}{81}+\frac{5\pi^2}{9}-\frac{4}{9}\zeta_3\right)
\nonumber\\[0.0em]
&\quad 
+ C_F^2\left(\frac{\pi^4}{72}\right)
+ C_F^2 \; \Delta \calC_{\delta}^{C_F}(R)
+ C_F C_A \; \Delta \calC_{\delta}^{C_A}(R)
+ C_F T_F n_f\; \Delta \calC_{\delta}^{n_f}(R) \,,
\label{eq:Cdelta:QQ:final}
\end{align}
which were derived using the semi-analytical method from Sec.~\ref{sec:analytical}. Explicitly, the various terms in these expressions read
\begin{align}
  \Delta d_2^{C_F}(R) &= 
  -32\zeta_3 + \frac{8 \pi^2 R^2}{3}-2R^4\,, 
    \notag \\
  \Delta d_2^{C_A}(R) &= 
  \left(-\frac{524}{9} + \frac{16 \pi^2}{3} + \frac{176}{3} \ln 2 \right) \ln R + \frac{3220}{27} - \frac{44 \pi^2}{9} - \frac{560 }{9} \ln 2 - \frac{176 }{3}  \ln^2 2 - 16 \zeta_3
  \notag \\ 
  &\quad
  - 8.45 R^2 +0.723 R^4 - 0.0169 R^6 + 0.000402 R^8  - 2.62 \cdot 10^{-5} R^{10}\,, 
  \notag \\
  \Delta d_2^{n_f}(R) &= \left(\frac{184}{9} - \frac{64}{3} \ln 2\right) \ln R -\frac{1256}{27} + \frac{16 \pi^2}{9} + \frac{256}{9} \ln 2 + \frac{64}{3}  \ln^2 2  \notag \\
  &\quad 
  - 0.706 R^2 + 0.0142 R^4 - 0.00217 R^6 + 0.000135 R^8 - 1.19 \cdot 10^{-5} R^{10} \,,
  \label{eq:delta-d2:QQ}
\end{align}
for the anomaly exponent, and 
\begin{align}
  \label{eq:DistributionsQQ}
  \Delta\calC_{\delta}^{C_F}(R)  &= 
   4.81 R^2+0.614 R^4-0.0764 R^6-0.000709 R^8  -3.85 \cdot 10^{-5} R^{10}\,,
  \notag \\
    \Delta\calC_{\delta}^{C_A}(R)  &= \left(-\frac{1622}{27} + \frac{548}{9} \ln 2 + \frac{88}{3} \ln^2 2 + 8 \zeta_3\right) \ln R -19.0 
  \notag\\
  &\quad
  -0.209 R^2-0.135 R^4+0.0228 R^6-0.000576 R^8+0.0000398 R^{10}
  \notag \\
  \Delta\calC_{\delta}^{n_f}(R)  &= \left(\frac{652}{27} - \frac{232}{9} \ln 2 - \frac{32}{3} \ln^2 2\right) \ln R + 1.27 
  \notag\\
  &\quad
  + 0.0538 R^2 - 0.0209 R^4 + 0.000599 R^6 - 0.000146 R^8 + 9.93 \cdot 10^{-6} R^{10}\,,
\end{align}
for the coefficient of the delta function. We recall that both of these coefficients are constrained by Casimir scaling, and they are therefore related to the expressions in  \eqref{eq:delta-d2} and \eqref{eq:DistributionsGG}. The agreement between our numerical and semi-analytical results provides a check of our calculation.

We finally turn to the grid contributions, for which we use the same colour decomposition as in~\cite{Bell:2022nrj}, 
\begin{align}
{I}_{q\leftarrow q}^{(2,\grid)}(z,R)&=
C_F^2 \; {I}_{q\leftarrow q}^{(2,C_F)}(z)
+ C_F C_A \; {I}_{q\leftarrow q}^{(2,C_A)}(z)   
+ C_F T_F n_f \; {I}_{q\leftarrow q}^{(2,n_f)}(z)
+ C_F T_F \; {I}_{q\leftarrow q}^{(2,T_F)}(z)\,,
\nonumber\\
{I}_{q\leftarrow g}^{(2,\grid)}(z,R) &=
C_F T_F \; {I}_{q\leftarrow g}^{(2,C_F)}(z) 
+ C_A T_F \; {I}_{q\leftarrow g}^{(2,C_A)}(z) \,,
\nonumber\\
{I}_{q\leftarrow \bar q}^{(2,\grid)}(z,R) &=
C_F (C_A - 2 C_F) \; {I}_{q\leftarrow \bar q}^{(2,C_{AF})}(z)  
+ C_F T_F \; {I}_{q\leftarrow q}^{(2,T_F)}(z) \,,
\nonumber\\
{I}_{q\leftarrow q'}^{(2,\grid)}(z,R) &=
{I}_{q\leftarrow \bar q'}^{(2,\grid)}(z,R) =
C_F T_F \; {I}_{q\leftarrow q}^{(2,T_F)}(z)\,.
\label{eq:gridQuark}
\end{align}
In total there are thus seven independent kernels in this case, which implicitly depend on the jet radius $R$. Our results for these kernels are shown in Fig.~\ref{fig:QuarkNNLOkernels}, and they are also contained in the accompanying electronic file. We note that these numbers, which were obtained with our numerical setup, have sub-percent uncertainties, and they show a similar level of agreement in comparison with both our semi-analytical approach and the results from \cite{Abreu:2022zgo} as for the gluon channels.
 
\end{appendix}

\bibliography{pTveto}

\providecommand{\href}[2]{#2}\begingroup\raggedright\begin{thebibliography}{10}

\bibitem{Banfi:2012yh}
A.~Banfi, G.~P. Salam and G.~Zanderighi, \emph{{NLL+NNLO predictions for
  jet-veto efficiencies in Higgs-boson and Drell-Yan production}},
  \href{https://doi.org/10.1007/JHEP06(2012)159}{\emph{JHEP} {\bfseries 06}
  (2012) 159} [\href{https://arxiv.org/abs/1203.5773}{{\ttfamily 1203.5773}}].

\bibitem{Becher:2012qa}
T.~Becher and M.~Neubert, \emph{{Factorization and NNLL Resummation for Higgs
  Production with a Jet Veto}},
  \href{https://doi.org/10.1007/JHEP07(2012)108}{\emph{JHEP} {\bfseries 07}
  (2012) 108} [\href{https://arxiv.org/abs/1205.3806}{{\ttfamily 1205.3806}}].

\bibitem{Tackmann:2012bt}
F.~J. Tackmann, J.~R. Walsh and S.~Zuberi, \emph{{Resummation Properties of Jet
  Vetoes at the LHC}},
  \href{https://doi.org/10.1103/PhysRevD.86.053011}{\emph{Phys. Rev. D}
  {\bfseries 86} (2012) 053011}
  [\href{https://arxiv.org/abs/1206.4312}{{\ttfamily 1206.4312}}].

\bibitem{Banfi:2012jm}
A.~Banfi, P.~F. Monni, G.~P. Salam and G.~Zanderighi, \emph{{Higgs and Z-boson
  production with a jet veto}},
  \href{https://doi.org/10.1103/PhysRevLett.109.202001}{\emph{Phys. Rev. Lett.}
  {\bfseries 109} (2012) 202001}
  [\href{https://arxiv.org/abs/1206.4998}{{\ttfamily 1206.4998}}].

\bibitem{Becher:2013xia}
T.~Becher, M.~Neubert and L.~Rothen, \emph{{Factorization and
  $N^{3}LL_{p}$+NNLO predictions for the Higgs cross section with a jet veto}},
  \href{https://doi.org/10.1007/JHEP10(2013)125}{\emph{JHEP} {\bfseries 10}
  (2013) 125} [\href{https://arxiv.org/abs/1307.0025}{{\ttfamily 1307.0025}}].

\bibitem{Stewart:2013faa}
I.~W. Stewart, F.~J. Tackmann, J.~R. Walsh and S.~Zuberi, \emph{{Jet $p_T$
  resummation in Higgs production at $NNLL'+NNLO$}},
  \href{https://doi.org/10.1103/PhysRevD.89.054001}{\emph{Phys. Rev. D}
  {\bfseries 89} (2014) 054001}
  [\href{https://arxiv.org/abs/1307.1808}{{\ttfamily 1307.1808}}].

\bibitem{Banfi:2015pju}
A.~Banfi, F.~Caola, F.~A. Dreyer, P.~F. Monni, G.~P. Salam, G.~Zanderighi
  et~al., \emph{{Jet-vetoed Higgs cross section in gluon fusion at
  N$^{3}$LO+NNLL with small-$R$ resummation}},
  \href{https://doi.org/10.1007/JHEP04(2016)049}{\emph{JHEP} {\bfseries 04}
  (2016) 049} [\href{https://arxiv.org/abs/1511.02886}{{\ttfamily
  1511.02886}}].

\bibitem{Monni:2019yyr}
P.~F. Monni, L.~Rottoli and P.~Torrielli, \emph{{Higgs transverse momentum with
  a jet veto: a double-differential resummation}},
  \href{https://doi.org/10.1103/PhysRevLett.124.252001}{\emph{Phys. Rev. Lett.}
  {\bfseries 124} (2020) 252001}
  [\href{https://arxiv.org/abs/1909.04704}{{\ttfamily 1909.04704}}].

\bibitem{Shao:2013uba}
D.~Y. Shao, C.~S. Li and H.~T. Li, \emph{{Resummation Prediction on Higgs and
  Vector Boson Associated Production with a Jet Veto at the LHC}},
  \href{https://doi.org/10.1007/JHEP02(2014)117}{\emph{JHEP} {\bfseries 02}
  (2014) 117} [\href{https://arxiv.org/abs/1309.5015}{{\ttfamily 1309.5015}}].

\bibitem{Li:2014ria}
Y.~Li and X.~Liu, \emph{{High precision predictions for exclusive $VH$
  production at the LHC}},
  \href{https://doi.org/10.1007/JHEP06(2014)028}{\emph{JHEP} {\bfseries 06}
  (2014) 028} [\href{https://arxiv.org/abs/1401.2149}{{\ttfamily 1401.2149}}].

\bibitem{Jaiswal:2014yba}
P.~Jaiswal and T.~Okui, \emph{{Explanation of the $WW$ excess at the LHC by
  jet-veto resummation}},
  \href{https://doi.org/10.1103/PhysRevD.90.073009}{\emph{Phys. Rev. D}
  {\bfseries 90} (2014) 073009}
  [\href{https://arxiv.org/abs/1407.4537}{{\ttfamily 1407.4537}}].

\bibitem{Becher:2014aya}
T.~Becher, R.~Frederix, M.~Neubert and L.~Rothen, \emph{{Automated NNLL $+$ NLO
  resummation for jet-veto cross sections}},
  \href{https://doi.org/10.1140/epjc/s10052-015-3368-y}{\emph{Eur. Phys. J. C}
  {\bfseries 75} (2015) 154} [\href{https://arxiv.org/abs/1412.8408}{{\ttfamily
  1412.8408}}].

\bibitem{Wang:2015mvz}
Y.~Wang, C.~S. Li and Z.~L. Liu, \emph{{Resummation prediction on gauge boson
  pair production with a jet veto}},
  \href{https://doi.org/10.1103/PhysRevD.93.094020}{\emph{Phys. Rev. D}
  {\bfseries 93} (2016) 094020}
  [\href{https://arxiv.org/abs/1504.00509}{{\ttfamily 1504.00509}}].

\bibitem{Dawson:2016ysj}
S.~Dawson, P.~Jaiswal, Y.~Li, H.~Ramani and M.~Zeng, \emph{{Resummation of jet
  veto logarithms at N$^3$LL$_a$ + NNLO for $W^+ W^-$ production at the LHC}},
  \href{https://doi.org/10.1103/PhysRevD.94.114014}{\emph{Phys. Rev. D}
  {\bfseries 94} (2016) 114014}
  [\href{https://arxiv.org/abs/1606.01034}{{\ttfamily 1606.01034}}].

\bibitem{Campbell:2023cha}
J.~M. Campbell, R.~K. Ellis, T.~Neumann and S.~Seth, \emph{{Jet-veto
  resummation at N$^{3}$LL$_{p}$ + NNLO in boson production processes}},
  \href{https://doi.org/10.1007/JHEP04(2023)106}{\emph{JHEP} {\bfseries 04}
  (2023) 106} [\href{https://arxiv.org/abs/2301.11768}{{\ttfamily
  2301.11768}}].

\bibitem{Gavardi:2023aco}
A.~Gavardi, M.~A. Lim, S.~Alioli and F.~J. Tackmann, \emph{{NNLO+PS
  $W^{+}W^{-}$ production using jet veto resummation at NNLL$'$}},
  \href{https://doi.org/10.1007/JHEP12(2023)069}{\emph{JHEP} {\bfseries 12}
  (2023) 069} [\href{https://arxiv.org/abs/2308.11577}{{\ttfamily
  2308.11577}}].

\bibitem{Tackmann:2016jyb}
F.~J. Tackmann, W.~J. Waalewijn and L.~Zeune, \emph{{Impact of Jet Veto
  Resummation on Slepton Searches}},
  \href{https://doi.org/10.1007/JHEP07(2016)119}{\emph{JHEP} {\bfseries 07}
  (2016) 119} [\href{https://arxiv.org/abs/1603.03052}{{\ttfamily
  1603.03052}}].

\bibitem{Ebert:2016idf}
M.~A. Ebert, S.~Liebler, I.~Moult, I.~W. Stewart, F.~J. Tackmann, K.~Tackmann
  et~al., \emph{{Exploiting jet binning to identify the initial state of
  high-mass resonances}},
  \href{https://doi.org/10.1103/PhysRevD.94.051901}{\emph{Phys. Rev. D}
  {\bfseries 94} (2016) 051901}
  [\href{https://arxiv.org/abs/1605.06114}{{\ttfamily 1605.06114}}].

\bibitem{Fuks:2017vtl}
B.~Fuks and R.~Ruiz, \emph{{A comprehensive framework for studying $W'$ and
  $Z'$ bosons at hadron colliders with automated jet veto resummation}},
  \href{https://doi.org/10.1007/JHEP05(2017)032}{\emph{JHEP} {\bfseries 05}
  (2017) 032} [\href{https://arxiv.org/abs/1701.05263}{{\ttfamily
  1701.05263}}].

\bibitem{Arpino:2019fmo}
L.~Arpino, A.~Banfi, S.~J\"ager and N.~Kauer, \emph{{BSM $WW$ production with a
  jet veto}}, \href{https://doi.org/10.1007/JHEP08(2019)076}{\emph{JHEP}
  {\bfseries 08} (2019) 076}
  [\href{https://arxiv.org/abs/1905.06646}{{\ttfamily 1905.06646}}].

\bibitem{Bauer:2000yr}
C.~W. Bauer, S.~Fleming, D.~Pirjol and I.~W. Stewart, \emph{{An Effective field
  theory for collinear and soft gluons: Heavy to light decays}},
  \href{https://doi.org/10.1103/PhysRevD.63.114020}{\emph{Phys. Rev. D}
  {\bfseries 63} (2001) 114020}
  [\href{https://arxiv.org/abs/hep-ph/0011336}{{\ttfamily hep-ph/0011336}}].

\bibitem{Bauer:2001yt}
C.~W. Bauer, D.~Pirjol and I.~W. Stewart, \emph{{Soft collinear factorization
  in effective field theory}},
  \href{https://doi.org/10.1103/PhysRevD.65.054022}{\emph{Phys. Rev. D}
  {\bfseries 65} (2002) 054022}
  [\href{https://arxiv.org/abs/hep-ph/0109045}{{\ttfamily hep-ph/0109045}}].

\bibitem{Beneke:2002ph}
M.~Beneke, A.~P. Chapovsky, M.~Diehl and T.~Feldmann, \emph{{Soft collinear
  effective theory and heavy to light currents beyond leading power}},
  \href{https://doi.org/10.1016/S0550-3213(02)00687-9}{\emph{Nucl. Phys. B}
  {\bfseries 643} (2002) 431}
  [\href{https://arxiv.org/abs/hep-ph/0206152}{{\ttfamily hep-ph/0206152}}].

\bibitem{Becher:2010tm}
T.~Becher and M.~Neubert, \emph{{Drell-Yan Production at Small $q_T$,
  Transverse Parton Distributions and the Collinear Anomaly}},
  \href{https://doi.org/10.1140/epjc/s10052-011-1665-7}{\emph{Eur. Phys. J. C}
  {\bfseries 71} (2011) 1665}
  [\href{https://arxiv.org/abs/1007.4005}{{\ttfamily 1007.4005}}].

\bibitem{Becher:2011pf}
T.~Becher, G.~Bell and M.~Neubert, \emph{{Factorization and Resummation for Jet
  Broadening}},
  \href{https://doi.org/10.1016/j.physletb.2011.09.005}{\emph{Phys. Lett. B}
  {\bfseries 704} (2011) 276}
  [\href{https://arxiv.org/abs/1104.4108}{{\ttfamily 1104.4108}}].

\bibitem{Chiu:2012ir}
J.-Y. Chiu, A.~Jain, D.~Neill and I.~Z. Rothstein, \emph{{A Formalism for the
  Systematic Treatment of Rapidity Logarithms in Quantum Field Theory}},
  \href{https://doi.org/10.1007/JHEP05(2012)084}{\emph{JHEP} {\bfseries 05}
  (2012) 084} [\href{https://arxiv.org/abs/1202.0814}{{\ttfamily 1202.0814}}].

\bibitem{Bell:2020yzz}
G.~Bell, R.~Rahn and J.~Talbert, \emph{{Generic dijet soft functions at
  two-loop order: uncorrelated emissions}},
  \href{https://doi.org/10.1007/JHEP09(2020)015}{\emph{JHEP} {\bfseries 09}
  (2020) 015} [\href{https://arxiv.org/abs/2004.08396}{{\ttfamily
  2004.08396}}].

\bibitem{Abreu:2022sdc}
S.~Abreu, J.~R. Gaunt, P.~F. Monni and R.~Szafron, \emph{{The analytic two-loop
  soft function for leading-jet p$_{T}$}},
  \href{https://doi.org/10.1007/JHEP08(2022)268}{\emph{JHEP} {\bfseries 08}
  (2022) 268} [\href{https://arxiv.org/abs/2204.02987}{{\ttfamily
  2204.02987}}].

\bibitem{Bell:2022nrj}
G.~Bell, K.~Brune, G.~Das and M.~Wald, \emph{{The NNLO quark beam function for
  jet-veto resummation}},
  \href{https://doi.org/10.1007/JHEP01(2023)083}{\emph{JHEP} {\bfseries 01}
  (2023) 083} [\href{https://arxiv.org/abs/2207.05578}{{\ttfamily
  2207.05578}}].

\bibitem{Abreu:2022zgo}
S.~Abreu, J.~R. Gaunt, P.~F. Monni, L.~Rottoli and R.~Szafron, \emph{{Quark and
  gluon two-loop beam functions for leading-jet p$_{T}$ and slicing at NNLO}},
  \href{https://doi.org/10.1007/JHEP04(2023)127}{\emph{JHEP} {\bfseries 04}
  (2023) 127} [\href{https://arxiv.org/abs/2207.07037}{{\ttfamily
  2207.07037}}].

\bibitem{Bell:2018vaa}
G.~Bell, R.~Rahn and J.~Talbert, \emph{{Two-loop anomalous dimensions of
  generic dijet soft functions}},
  \href{https://doi.org/10.1016/j.nuclphysb.2018.09.026}{\emph{Nucl. Phys. B}
  {\bfseries 936} (2018) 520}
  [\href{https://arxiv.org/abs/1805.12414}{{\ttfamily 1805.12414}}].

\bibitem{Bell:2018oqa}
G.~Bell, R.~Rahn and J.~Talbert, \emph{{Generic dijet soft functions at
  two-loop order: correlated emissions}},
  \href{https://doi.org/10.1007/JHEP07(2019)101}{\emph{JHEP} {\bfseries 07}
  (2019) 101} [\href{https://arxiv.org/abs/1812.08690}{{\ttfamily
  1812.08690}}].

\bibitem{Bell:2023yso}
G.~Bell, B.~Dehnadi, T.~Mohrmann and R.~Rahn, \emph{{The NNLO soft function for
  N-jettiness in hadronic collisions}},
  \href{https://arxiv.org/abs/2312.11626}{{\ttfamily 2312.11626}}.

\bibitem{Bell:2021dpb}
G.~Bell, K.~Brune, G.~Das and M.~Wald, \emph{{Automation of Beam and Jet
  functions at NNLO}},
  \href{https://doi.org/10.21468/SciPostPhysProc.7.021}{\emph{SciPost Phys.
  Proc.} {\bfseries 7} (2022) 021}
  [\href{https://arxiv.org/abs/2110.04804}{{\ttfamily 2110.04804}}].

\bibitem{Bell:2022tmi}
G.~Bell, K.~Brune, G.~Das and M.~Wald, \emph{{Automated Calculation of Beam
  Functions at NNLO}}, \href{https://doi.org/10.22323/1.416.0026}{\emph{PoS}
  {\bfseries LL2022} (2022) 026}
  [\href{https://arxiv.org/abs/2208.04847}{{\ttfamily 2208.04847}}].

\bibitem{Wald:thesis}
M.~Wald, \emph{{Factorisation: Applications in collider and flavour physics}},
  Ph.D. thesis, University of Siegen, 2024.
\newblock
  \href{http://dx.doi.org/10.25819/ubsi/10468}{http://dx.doi.org/10.25819/ubsi/10468}.

\bibitem{Catani:2013tia}
S.~Catani, L.~Cieri, D.~de~Florian, G.~Ferrera and M.~Grazzini,
  \emph{{Universality of transverse-momentum resummation and hard factors at
  the NNLO}},
  \href{https://doi.org/10.1016/j.nuclphysb.2014.02.011}{\emph{Nucl. Phys. B}
  {\bfseries 881} (2014) 414}
  [\href{https://arxiv.org/abs/1311.1654}{{\ttfamily 1311.1654}}].

\bibitem{Gehrmann:2014yya}
T.~Gehrmann, T.~Luebbert and L.~L. Yang, \emph{{Calculation of the transverse
  parton distribution functions at next-to-next-to-leading order}},
  \href{https://doi.org/10.1007/JHEP06(2014)155}{\emph{JHEP} {\bfseries 06}
  (2014) 155} [\href{https://arxiv.org/abs/1403.6451}{{\ttfamily 1403.6451}}].

\bibitem{Luo:2019szz}
M.-x. Luo, T.-Z. Yang, H.~X. Zhu and Y.~J. Zhu, \emph{{Quark Transverse Parton
  Distribution at the Next-to-Next-to-Next-to-Leading Order}},
  \href{https://doi.org/10.1103/PhysRevLett.124.092001}{\emph{Phys. Rev. Lett.}
  {\bfseries 124} (2020) 092001}
  [\href{https://arxiv.org/abs/1912.05778}{{\ttfamily 1912.05778}}].

\bibitem{Ebert:2020yqt}
M.~A. Ebert, B.~Mistlberger and G.~Vita, \emph{{Transverse momentum dependent
  PDFs at N$^3$LO}}, \href{https://doi.org/10.1007/JHEP09(2020)146}{\emph{JHEP}
  {\bfseries 09} (2020) 146}
  [\href{https://arxiv.org/abs/2006.05329}{{\ttfamily 2006.05329}}].

\bibitem{Luo:2020epw}
M.-x. Luo, T.-Z. Yang, H.~X. Zhu and Y.~J. Zhu, \emph{{Unpolarized quark and
  gluon TMD PDFs and FFs at N$^{3}$LO}},
  \href{https://doi.org/10.1007/JHEP06(2021)115}{\emph{JHEP} {\bfseries 06}
  (2021) 115} [\href{https://arxiv.org/abs/2012.03256}{{\ttfamily
  2012.03256}}].

\bibitem{Becher:2011dz}
T.~Becher and G.~Bell, \emph{{Analytic Regularization in Soft-Collinear
  Effective Theory}},
  \href{https://doi.org/10.1016/j.physletb.2012.05.016}{\emph{Phys. Lett. B}
  {\bfseries 713} (2012) 41} [\href{https://arxiv.org/abs/1112.3907}{{\ttfamily
  1112.3907}}].

\bibitem{Ritzmann:2014mka}
M.~Ritzmann and W.~J. Waalewijn, \emph{{Fragmentation in Jets at NNLO}},
  \href{https://doi.org/10.1103/PhysRevD.90.054029}{\emph{Phys. Rev. D}
  {\bfseries 90} (2014) 054029}
  [\href{https://arxiv.org/abs/1407.3272}{{\ttfamily 1407.3272}}].

\bibitem{Kosower:1999rx}
D.~A. Kosower and P.~Uwer, \emph{{One loop splitting amplitudes in gauge
  theory}}, \href{https://doi.org/10.1016/S0550-3213(99)00583-0}{\emph{Nucl.
  Phys. B} {\bfseries 563} (1999) 477}
  [\href{https://arxiv.org/abs/hep-ph/9903515}{{\ttfamily hep-ph/9903515}}].

\bibitem{Bern:1999ry}
Z.~Bern, V.~Del~Duca, W.~B. Kilgore and C.~R. Schmidt, \emph{{The infrared
  behavior of one loop QCD amplitudes at next-to-next-to leading order}},
  \href{https://doi.org/10.1103/PhysRevD.60.116001}{\emph{Phys. Rev. D}
  {\bfseries 60} (1999) 116001}
  [\href{https://arxiv.org/abs/hep-ph/9903516}{{\ttfamily hep-ph/9903516}}].

\bibitem{Sborlini:2013jba}
G.~F.~R. Sborlini, D.~de~Florian and G.~Rodrigo, \emph{{Double collinear
  splitting amplitudes at next-to-leading order}},
  \href{https://doi.org/10.1007/JHEP01(2014)018}{\emph{JHEP} {\bfseries 01}
  (2014) 018} [\href{https://arxiv.org/abs/1310.6841}{{\ttfamily 1310.6841}}].

\bibitem{Campbell:1997hg}
J.~M. Campbell and E.~W.~N. Glover, \emph{{Double unresolved approximations to
  multiparton scattering amplitudes}},
  \href{https://doi.org/10.1016/S0550-3213(98)00295-8}{\emph{Nucl. Phys. B}
  {\bfseries 527} (1998) 264}
  [\href{https://arxiv.org/abs/hep-ph/9710255}{{\ttfamily hep-ph/9710255}}].

\bibitem{Catani:1998nv}
S.~Catani and M.~Grazzini, \emph{{Collinear factorization and splitting
  functions for next-to-next-to-leading order QCD calculations}},
  \href{https://doi.org/10.1016/S0370-2693(98)01513-5}{\emph{Phys. Lett. B}
  {\bfseries 446} (1999) 143}
  [\href{https://arxiv.org/abs/hep-ph/9810389}{{\ttfamily hep-ph/9810389}}].

\bibitem{Borowka:2017idc}
S.~Borowka, G.~Heinrich, S.~Jahn, S.~P. Jones, M.~Kerner, J.~Schlenk et~al.,
  \emph{{pySecDec: a toolbox for the numerical evaluation of multi-scale
  integrals}}, \href{https://doi.org/10.1016/j.cpc.2017.09.015}{\emph{Comput.
  Phys. Commun.} {\bfseries 222} (2018) 313}
  [\href{https://arxiv.org/abs/1703.09692}{{\ttfamily 1703.09692}}].

\bibitem{Hahn:2004fe}
T.~Hahn, \emph{{CUBA: A Library for multidimensional numerical integration}},
  \href{https://doi.org/10.1016/j.cpc.2005.01.010}{\emph{Comput. Phys. Commun.}
  {\bfseries 168} (2005) 78}
  [\href{https://arxiv.org/abs/hep-ph/0404043}{{\ttfamily hep-ph/0404043}}].

\bibitem{Becher:2012qc}
T.~Becher and G.~Bell, \emph{{NNLL Resummation for Jet Broadening}},
  \href{https://doi.org/10.1007/JHEP11(2012)126}{\emph{JHEP} {\bfseries 11}
  (2012) 126} [\href{https://arxiv.org/abs/1210.0580}{{\ttfamily 1210.0580}}].

\bibitem{Kang:2022nft}
Z.-B. Kang, K.~Samanta, D.~Y. Shao and Y.-L. Zeng, \emph{{Transverse momentum
  dependent distribution functions in the threshold limit}},
  \href{https://doi.org/10.1007/JHEP11(2023)220}{\emph{JHEP} {\bfseries 11}
  (2023) 220} [\href{https://arxiv.org/abs/2211.08341}{{\ttfamily
  2211.08341}}].

\bibitem{Becher:2006mr}
T.~Becher, M.~Neubert and B.~D. Pecjak, \emph{{Factorization and Momentum-Space
  Resummation in Deep-Inelastic Scattering}},
  \href{https://doi.org/10.1088/1126-6708/2007/01/076}{\emph{JHEP} {\bfseries
  01} (2007) 076} [\href{https://arxiv.org/abs/hep-ph/0607228}{{\ttfamily
  hep-ph/0607228}}].

\bibitem{Becher:2009qa}
T.~Becher and M.~Neubert, \emph{{On the Structure of Infrared Singularities of
  Gauge-Theory Amplitudes}},
  \href{https://doi.org/10.1088/1126-6708/2009/06/081}{\emph{JHEP} {\bfseries
  06} (2009) 081} [\href{https://arxiv.org/abs/0903.1126}{{\ttfamily
  0903.1126}}].

\bibitem{Curci:1980uw}
G.~Curci, W.~Furmanski and R.~Petronzio, \emph{{Evolution of Parton Densities
  Beyond Leading Order: The Nonsinglet Case}},
  \href{https://doi.org/10.1016/0550-3213(80)90003-6}{\emph{Nucl. Phys. B}
  {\bfseries 175} (1980) 27}.

\bibitem{Furmanski:1980cm}
W.~Furmanski and R.~Petronzio, \emph{{Singlet Parton Densities Beyond Leading
  Order}}, \href{https://doi.org/10.1016/0370-2693(80)90636-X}{\emph{Phys.
  Lett. B} {\bfseries 97} (1980) 437}.

\bibitem{Ellis:1996nn}
R.~K. Ellis and W.~Vogelsang, \emph{{The Evolution of parton distributions
  beyond leading order: The Singlet case}},
  \href{https://arxiv.org/abs/hep-ph/9602356}{{\ttfamily hep-ph/9602356}}.

\end{thebibliography}\endgroup

\end{document}